\documentclass[preprint]{aastex63}

\usepackage{textcomp}
\usepackage{graphicx}
\usepackage{natbib}

\shorttitle{Tilting Uranus: Collisions versus Spin--Orbit Resonance}
\shortauthors{Rogoszinski and Hamilton}

\begin{document}

\title{Tilting Uranus: Collisions versus Spin--Orbit Resonance}

\author{Zeeve Rogoszinski}
\affiliation{Astronomy Department  \\
	University of Maryland \\
	College Park, MD 20742, USA}

\author{Douglas P. Hamilton}
\affiliation{Astronomy Department  \\
	University of Maryland \\
	College Park, MD 20742, USA}

\email{zero@umd.edu, dphamil@umd.edu}

\begin{abstract}
	
	In this paper, we investigate whether Uranus's 98\textdegree~obliquity was a by-product of a secular spin-orbit resonance assuming that the planet originated closer to the Sun. In this position, Uranus's spin precession frequency is fast enough to resonate with another planet located beyond Saturn. Using numerical integration, we show that resonance capture is possible in a variety of past solar system configurations, but that the timescale required to tilt the planet to 90\textdegree~is of the order $\sim\!10^{8}$ yr---a timespan that is uncomfortably long. A resonance kick could tilt the planet to a significant 40\textdegree~in $\sim\!10^{7}$ yr only if conditions were ideal. We also revisit the collisional hypothesis for the origin of Uranus's large obliquity. We consider multiple impacts with a new collisional code that builds up a planet by summing the angular momentum imparted from impactors. Because gas accretion imparts an unknown but likely large part of the planet's spin angular momentum, we compare different collisional models for tilted, untilted, spinning, and nonspinning planets. We find that a 1 $M_{\oplus}$ strike is sufficient to explain the planet's current spin state, but that two $0.5\,M_{\oplus}$ collisions produce better likelihoods. Finally, we investigate hybrid models and show that resonances must produce a tilt of at least $\sim\!40$\textdegree~for any noticeable improvements to the collision model. Because it is difficult for spin-orbit resonances to drive Uranus's obliquity to 98\textdegree~even under these ideal conditions, giant impacts seem inescapable. 
	

\end{abstract}

\section{Introduction}

Uranus's 98\textdegree~obliquity, the angle between the planet's spin axis and normal to its orbital plane, is perhaps the most unusual feature in our solar system. The most accepted explanation for its origin is a giant collision with an Earth-sized object that struck Uranus at polar latitudes during the late stages of planetary formation \citep{1982AREPS..10...61H,1989Metic..24R.251B,1990Icar...84..528K,1992Icar...99..167S,1997P&SS...45..181P,2012Icar..219..737M,2015A&A...582A..99I,2018ApJ...861...52K,2019MNRAS.487.5029K,2019AJ....157...13K,2020NatAs...4..880I,2020MNRAS.492.5336R}. Collisions between massive objects are an expected part of solar system formation; indeed, our own Moon was likely formed as a result of a collision between Earth and a Mars-sized object \citep{2001Natur.412..708C}. There are problems with a collisional origin of Uranus's obliquity though. These impacts could significantly alter the planet's primordial spin rate, yet both Uranus and Neptune spin at similar periods ($T_{U}=17.2$ hr,  $T_{N}=16.1$ hr). Just as with Jupiter and Saturn, the two ice giants likely acquired their nearly identical spin rates while accreting their massive gaseous atmospheres from a circumplanetary disk \citep{2018AJ....155..178B, 2018NatAs...2..138B}. 

Additionally, \cite{2012Icar..219..737M} argue that for Uranus's regular satellites to orbit prograde around the planet, two or more collisions would be necessary. Tilting from 0\textdegree~to 98\textdegree~with a single impact would lead to nodal precession of the satellites, and they would form a torus around the tilted spin axis. The satellites would then cross orbits and undergo mutual collisions, and this collision damping would allow the satellites to eventually realign with the planet's equatorial plane; however, the resulting protosatellite disk would preserve its pre-impact angular momentum and hence would form retrograde satellites. Maintaining the orientation of its regular satellites is possible if Uranus's initial obliquity was large so that the final impact tilts the planet by less than 90\textdegree.

\cite{2020NatAs...4..880I} circumvent this multicollision issue by suggesting that the Uranian satellite system was a by-product of debris from an ice-rich giant impact. Previous simulations showed that a debris disk from a single rocky impactor would generate a disk 100 times more massive and 10 times smaller in size than the current Uranian satellite system \citep{1992Icar...99..167S,2018ApJ...861...52K,2019MNRAS.487.5029K,2019AJ....157...13K,2020MNRAS.492.5336R}. But, an icy impactor would eject a water-vapor-rich disk that viscously evolves until particles recondense to ice. Nearly all of the debris falls back onto Uranus, but the remaining 1\% of the disk spreads to 10 times the size of the initial debris disk and forms the equatorial satellite system observed today; alternatively, having the planet already tilted beyond 30\textdegree~would also reduce the number of impacts required by \cite{2012Icar..219..737M} back to one. If the Uranian satellites were indeed formed from a circumplanetary disk \citep{2018ApJ...868L..13S} rather than a debris disk, then exciting Uranus's pre-impact tilt through some mechanism other than collisions is desirable.

In this paper, we will explore an alternative collisionless approach based on the resonant capture explanation for Saturn's 27\textdegree~obliquity. Because Saturn is composed of mostly hydrogen and helium gas \citep{2005ApJ...626L..57A,2005AREPS..33..493G}, we would expect gas accretion during planet formation to conserve angular momentum and force any primordial obliquity to $\epsilon\sim0$\textdegree. A collisional explanation would then require an impactor of $6-7.2\,M_{\oplus}$ \citep{parisi2002model}, which is even more unlikely than the putative Uranus strike. Juno mission observations of Jupiter's gravitational field suggest that the planet's core is diluted of heavy elements \citep{2017GeoRL..44.4649W,2019ApJ...872..100D}, and \cite{2019Natur.572..355L} posit that a $5\,M_{\oplus}$ impact can mix these metals within the planet's inner envelope. This model implies that such collisions could be common in the early solar system, but in situ explanations, such as erosion of the core from convective mixing \citep{2004jpsm.book...35G,2012ApJ...745...54W} or planetesimal enrichment \citep{2011MNRAS.416.1419H,2017ApJ...836..227L}, are still viable alternatives and do not require stochastic cataclysmic events. Instead, Saturn's obliquity can best be explained without collisions entirely by an ongoing secular spin-orbit resonance between the precession frequencies of Saturn's spin axis and Neptune's orbital pole \citep{2004AJ....128.2510H,2004AJ....128.2501W}. And even Jupiter's small tilt may have resulted from a resonance with either Uranus or Neptune \citep{2006ApJ...640L..91W,2015ApJ...806..143V}. A significant advantage of this model is that the gradual increase of Saturn's obliquity preserves both the planet's spin period and the orbits of its satellite system, which would eliminate all of the issues present in the giant impact hypothesis for Uranus \citep{1965AJ.....70....5G}.


Uranus's current spin precession frequency today is too slow to match any of the planets' orbital precession rates, but that may not have been the case in the past. \cite{2010ApJ...712L..44B} posit that a resonance is possible if Uranus harbored a moon large enough so that the planet's spin axis could precess sufficiently fast to resonate with its own orbit. This moon would, however, have to be larger than all known moons (between the mass of Ganymede and Mars), have to be located far from Uranus ($\approx50$ Uranian radii), and then have to disappear somehow, perhaps during planetary migration. 

A more promising solution is instead to place a circumplanetary disk of at least $4.5\times10^{-3}\,M_{\oplus}$ around Uranus during the last stage of its formation \citep{1982AREPS..10...61H,2020ApJ...888...60R}. Because Uranus must have harbored a massive circumplanetary disk to account for its gaseous atmosphere, and \cite{2018ApJ...868L..13S} calculated a circumplanetary disk of around $10^{-2}\,M_{\oplus}$, capturing into a spin-orbit resonance by linking Uranus's pole precession to its nodal precession seems plausible during formation. \cite{2020ApJ...888...60R} find that a 70\textdegree~kick is possible within the accretion time span of 1 Myr, and that while a subsequent impactor is still necessary to fully account for Uranus's 98\textdegree~tilt, it only needs to be 0.5 $M_{\oplus}$. The odds of this collision generating Uranus' current spin state are significantly greater, but to attain the 70\textdegree~obliquity, Uranus's orbital inclination would need to be around 10\textdegree. An inclination this high is a little uncomfortable and hints that further improvements to the model may be necessary. For instance, \cite{2018CeMDA.130...11Q} demonstrated a similar set of resonance arguments that are not sensitive to a planet's orbital inclination and that are capable of pushing a planet's obliquity beyond 90\textdegree. These arguments include mean-motion terms which arise naturally if the planets are configured in a resonance chain \citep{2019NatAs...3..424M}. 


Here, we investigate yet another possibility by placing Uranus closer to the Sun where tidal forces are stronger and precession timescales are shorter. This will require us to make some optimistic modifications to the planets' initial configurations in order to generate the desired resonance, as will be seen below. If our models yield fruitful results, then these assumptions will need to be carefully examined in the larger context of solar system formation. Furthermore, we also revisit the single and multicollision explanation in Section 4, as well as hybrid resonance and collision models. We will then critically compare all of these resonance and collisional models.

\section{Capture into a Secular Spin--Orbit Resonance} \label{rc}
\subsection{Initial Conditions}

Gravitational torques from the Sun on an oblate planet cause the planet's spin axis to precess backwards, or regress, about the normal to its orbital plane \citep{1966AJ.....71..891C}. Similarly, gravitational perturbations cause a planet's inclined orbit to regress around the Sun. A match between these two precession frequencies results in a secular spin-orbit resonance. In this case, the spin axis remains fixed relative to the planet's orbital pole, and the two vectors precess about the normal to the invariable plane. The longitudes of the two axial vectors, $\phi_{\alpha}$ and $\phi_{g}$, are measured from a reference polar direction to projections onto the invariable plane, and the resonance argument is given as \citep{2004AJ....128.2510H}
\begin{equation}\label{resarg}
\Psi = \phi_{\alpha} - \phi_{g}.
\end{equation}


The precession rate of Uranus's spin axis can be derived from first principles by considering the torques of the Sun and the Uranian moons on the planet's equatorial bulge. Following \cite{1966AJ.....71..891C}, if $\hat{\sigma}$ is a unit vector that points in the direction of the total angular momentum of the Uranian system, then
\begin{equation}\label{diffeq}
\frac{d\hat{\sigma}}{dt} = {\alpha}(\hat{\sigma}\times\hat{n})(\hat{\sigma}\cdot\hat{n})
\end{equation}
where $\hat{n}$ is a unit vector pointing in the direction of Uranus's orbital angular momentum, $\alpha$ is the spin precession rate near zero degree tilts, and $t$ is time. Uranus's axial precession period is therefore
\begin{equation}\label{period}
T_{\alpha}=\frac{2\pi}{{\alpha}\cos\epsilon},
\end{equation}
where $\epsilon$ is the obliquity and $\cos\epsilon = \hat{\sigma}\cdot\hat{n}$. The precession frequency near zero obliquity, $\alpha$, incorporates the torques from the Sun and the planet's moons on the central body \citep{1991Icar...89...85T}:
\begin{equation}\label{prec}
{\alpha} = \frac{3{n}^{2}}{2} \frac{{J}_{2}(1-\frac{3}{2}\sin^{2}{\theta}_{p}) + q}{K\omega \cos\theta_{p} + l}.
\end{equation}

\noindent Here, $n = (GM_{\odot}/r_{p}^{3})^{1/2}$ is the orbital angular speed of the planet, $G$ is the gravitational constant, $M_{\odot}$ is the Sun's mass, $r_{p}$ is the Sun--planet distance, $\omega$ is the planet's spin angular speed, $J_{2}$ is its quadrupole gravitational moment, and $K$ is its moment of inertia normalized by $M_{p}R_{p}^{2}$. For Uranus today, $M_{p}=14.5\,M_{\oplus}$, $R_{p}=2.56\times 10^{9}$ cm, $K=0.225$, and $J_{2}=0.00334343$.\footnote{All physical values of the solar system are courtesy of NASA Goddard Space Flight Center: http://nssdc.gsfc.nasa.gov/planetary/factsheet/} The parameter
$
q\equiv\frac{1}{2} {\sum}_{i} ({M_{i}}/{M_{p}})({a_{i}}/{R_{p}})^{2}(1-\frac{3}{2}\sin^{2}{\theta}_{i})
$
is the effective quadrupole coefficient of the satellite system, and
$
l\equiv{R}_{p}^{-2} {\sum}_{i} ({M_{i}}/{M_{p}})(GM_{p}a_{i})^{\frac{1}{2}}\cos\theta_{i} 
$
is the angular momentum of the satellite system divided by $M_{p}R_{p}^{2}$. The masses and semi-major axes of the satellites are $M_{i}$ and $a_{i}$, $\cos\theta_{p} = \hat{s}\cdot\hat{\sigma}$, and $\cos\theta_{i} = \hat{l_{i}}\cdot\hat{\sigma}$, where $\hat{s}$ is the direction of the spin angular momenta of the central body, and $\hat{l_{i}}$ is the normal to the satellite's orbit \citep{1991Icar...89...85T}. Note that $M_{i}\ll{M}_{p}$, where $M_{p}$ is the mass of the planet, and because the satellite orbits are nearly equatorial, we can take $\theta_{p} = \theta_{i} = 0$.  

Torques from the main Uranian satellites on the planet contribute significantly to its precessional motion, while those from other planets and satellites can be ignored. We therefore limit ourselves to Uranus's major moons---Oberon, Titania, Umbriel, Ariel, and Miranda. We find $q=0.01558$ which is about 4.7 times larger than Uranus's $J_{2}$, and $l=2.41\times10^{-7}$, which is smaller than $K\omega$ by about a factor of 100. So from Equation \ref{prec}, the effective quadrupole coefficient of the satellite system plays a much more significant role in the planet's precession period than the angular momentum of the satellite system. At its current obliquity, $\epsilon=98$\textdegree, Uranus's precession period is about 210 million years (or $\alpha=0.0062$ arcsec yr$^{-1}$), and reducing Uranus's obliquity to 0\textdegree~results in a precession period 7.2 times faster: 29 million years (or $\alpha=0.045$ arcsec yr$^{-1}$). This pole precession rate is much longer than any of the giant planets' fundamental frequencies \citep{1999ssd..book.....M}, but it can be sped up to $\approx\,2$ Myr by placing Uranus at around 7 au. This is just fast enough for Uranus to resonate with a similar planet---Neptune---located beyond Saturn.

Placing Uranus's orbit between those of Jupiter and Saturn is not entirely ad hoc. \cite{1999Natur.402..635T,2002AJ....123.2862T,2003Icar..161..431T} argue that at least the ice giants' cores might have formed between Jupiter and Saturn (4--10 au), as the timescales there for the accretion of planetesimals through an oligarchic growth model, when the large bodies in the planetary disk dominate the accretion of surrounding planetesimals, are more favorable than farther away. The Nice model \citep{2005Natur.435..466G, 2005Natur.435..462M, 2005Natur.435..459T} places Uranus closer to the Sun but beyond Saturn for similar reasons; however, having the ice giants form between Jupiter and Saturn is not inconsistent with the Nice model. If Uranus and Neptune were indeed formed between Jupiter and Saturn and later ejected sequentially, then a secular spin-orbit resonance between Uranus and Neptune is possible. Note that such close encounters would not yield any significant obliquity excitations because the perturbing torque is too weak as it depends on the planet's gravitational quadrupole moment \citep{2007Icar..190..103L}. A related possibility that is also sufficient for our purposes is if the planets were formed from pebble accretion, as the pebble isolation mass can be similar everywhere in the outer solar system \citep{2014A&A...572A..35L}, allowing Neptune to be initially formed beyond Saturn and Uranus between Jupiter and Saturn. In the following, we assume that Uranus is fully formed with its satellites located near their current configurations to derive the spin axis precession rate. We also include only the known giant planets because adding a third or fourth ice giant, as suggested by the Nice model to better reproduce the solar system \citep{2011ApJ...742L..22N,2012ApJ...744L...3B,2012AJ....144..117N}, would increase the planet's orbital precession rates and make it more difficult for Uranus to obtain a spin-orbit resonance. Only if we succeed to tilt Uranus reasonably under these conditions would we consider introducing more giant planets to the model.

\subsection{Method}

\begin{figure}[h]
	\centering
	\includegraphics[width=0.7\textwidth]{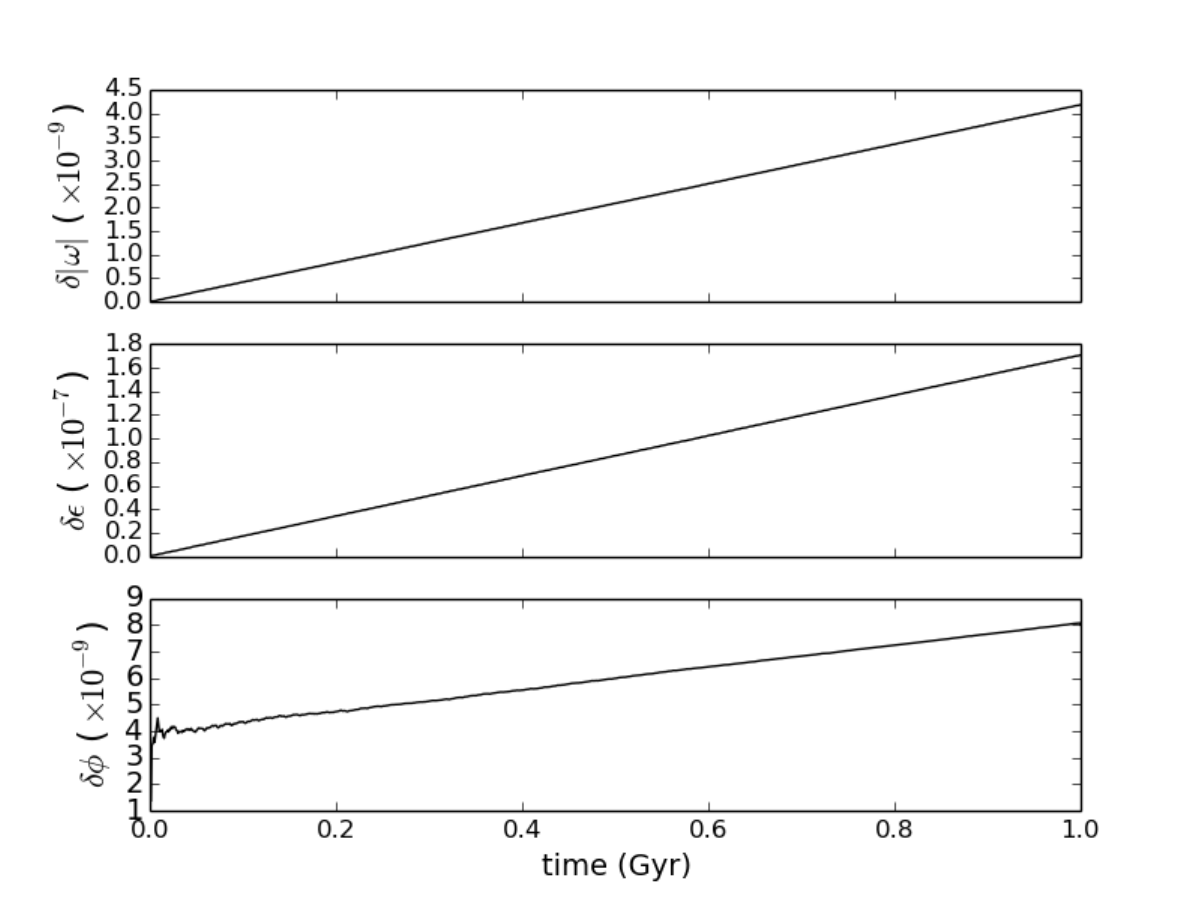}
	\caption{The calculated relative error of three quantities describing Uranus's spin axis. Here $\omega$ is the unit vector pointing in the direction of Uranus's spin axis. $\epsilon$ is the planet's obliquity and $\phi$ is the planet's spin longitude of the ascending node. All quantities should be constant with time as the system only contains the Sun and Uranus. Numerical errors at the levels shown here are sufficiently low for our purposes. }
	\label{plot}
\end{figure}

Calculating Uranus's obliquity evolution requires tracking the planets' orbits while also appropriately tuning Neptune's nodal precession rate. We use the HNBody Symplectic Integration package \citep{2002DDA....33.0802R} to track the motion of bodies orbiting a central massive object using symplectic integration techniques based on two-body Keplerian motion, and we move Neptune radially with an artificial drag force oriented along the velocity vector using the package HNDrag. These packages do not follow spins, so we have written an integrator that uses a fifth-order Runge-Kutta algorithm \citep{1992nrca.book.....P} and reads in HNBody data to calculate Uranus's axial orientation due to torques applied from the Sun (Equation \ref{diffeq}). For every time step, the integrator requires the distance between the Sun and Uranus. Because HNBody outputs the positions and velocities at a given time frequency different from the adaptive step that our precession integrator uses, calculating the precessional motion requires interpolation. To minimize interpolation errors, we use a torque averaged over an orbital period that is proportional to $\langle r_{p}^{-3}\rangle=a_{p}^{-3}(1-e_{p}^{2})^{-\frac{3}{2}}$, where $a_{p}$ is the planet's semi-major axis and $e_{p}$ is its eccentricity. This is an excellent approximation because Uranus's orbital period is $10^{5}-10^{6}$ times shorter than its precession period. We tested the code for a two-body system consisting of just the Sun and Uranus, and recovered the analytic result for the precession of the spin axis (Figure \ref{plot}).

For our simulations, we place Jupiter and Saturn near their current locations (5 au and 9 au, respectively), Uranus at 7 au, and Neptune well beyond Saturn at 17 au. Leaving Uranus in between the two gas giants for more than a few million years is unstable \citep{1973Icar...20..422L,1989Icar...79..223F,1990AJ....100.1680G,1993AJ....105.1987H}, but eccentricity dampening from remnant planetesimals can delay the instability. Scattering between Uranus and the planetesimals provides a dissipative force that temporarily prevents Uranus from being ejected, and we mimic this effect by applying an artificial force to damp Uranus's eccentricity. We apply the force in the orbital plane and perpendicular to the orbital velocity to damp the eccentricity while preventing changes to the semi-major axis \citep{1992fcm..book.....D}. With Uranus's orbit relatively stable, we then seek a secular resonance between its spin and Neptune's orbit. 

\subsection{A Secular Resonance}

Capturing into a spin-orbit resonance also requires the two angular momentum vectors, the planet's spin axis and an orbital pole, and the normal to the invariable plane be coplanar. Equilibria about which the resonance angle librates are called ``Cassini states'' \citep{1966AJ.....71..891C,1969AJ.....74..483P,1975AJ.....80...64W,2004AJ....128.2501W}, and there are multiple vector orientations that can yield a spin-orbit resonance. In our case, the resonance angle, $\Psi$, librates about Cassini state 2 because Uranus's spin axis and Neptune's orbital pole precess on opposite sides of the normal to the invariable plane. 

As Neptune migrates outwards away from the Sun, its nodal precession frequency slows until a resonance is reached with Uranus's spin precession rate. If the consequence of the resonance is that Uranus's obliquity increases \citep{1974JGR....79.3375W}, then its spin precession period increases as well (Equation (\ref{period})) and the resonance can persist. The time evolution of the resonance angle and obliquity are given by \cite{2004AJ....128.2510H}:
\begin{equation}\label{resdt}
\dot{\Psi}=-\alpha\cos\epsilon - g\cos{I}
\end{equation}
\begin{equation}\label{obldt}
\dot{\epsilon}=g\sin{I}\sin{\Psi}
\end{equation}
where $g$ is the negative nodal precession rate, and $I$ is the amplitude of the inclination induced by Neptune's perturbation on Uranus's orbit. If Neptune migrates outward slowly enough, then $\dot{\Psi}$ is small and the two planets can remain in resonance nearly indefinitely. 


\begin{figure}[h]
	\centering
	\includegraphics[width=0.7\textwidth]{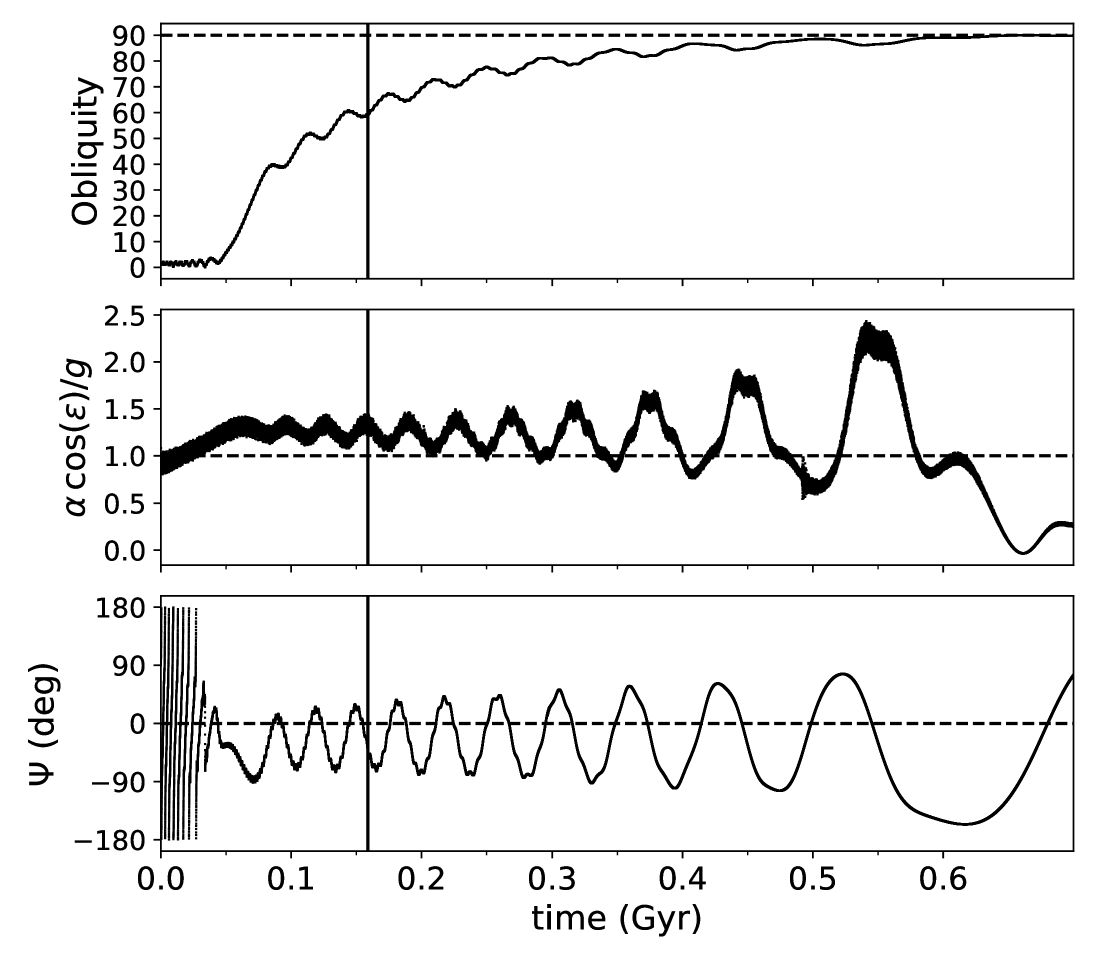}
	\caption{A resonance capture. The top panel shows Uranus's obliquity evolution over time. The middle panel shows the evolution of the precession frequencies with the dashed line indicating the resonance location, and the bottom panel shows the resonance angle ($\Psi$). The solid vertical line at $t\approx 150$ Myr indicates when Neptune reaches it current location at 30 au. In this simulation, resonance is established at $t=0.05$ Gyr when Neptune is at $\approx 24$ au, and it breaks at $t>0.7$ Gyr with Neptune near $\approx 120$ au. Stopping Neptune at 30 au, we find that this capture could account for perhaps half of Uranus's extreme tilt. Here, Uranus is located at $a_{U}=7$ au, with its current equatorial radius. Neptune's inclination is set to twice its current value at $i_{N}=4$\textdegree, which strengthens the resonance.}
	\label{capture}
\end{figure}

\begin{figure}[h]
	\centering
	\includegraphics[width=0.7\textwidth]{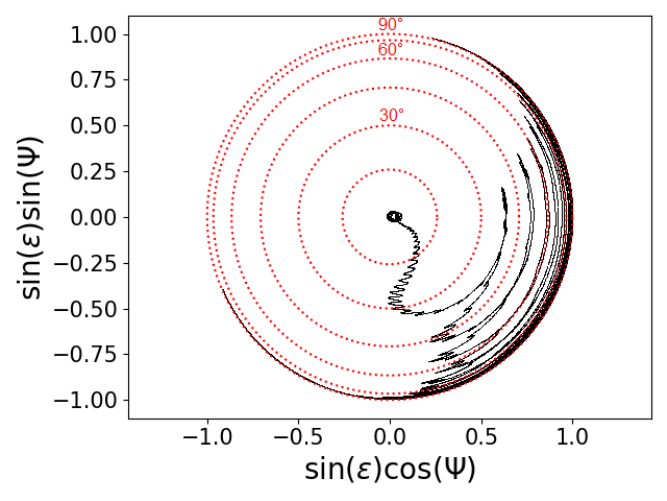}
	\caption{The corresponding polar plot to Figure \ref{capture} where Neptune is migrating well within the adiabatic limit. The short-period oscillations here are at the pole precession rate while the longer oscillations are the librations about the equilibrium point, which itself is moving to higher obliquities (to the right). The red dotted circles represent points of constant obliquity in increments of 15\textdegree.}
	\label{polar:capture}
\end{figure}

Figure \ref{capture} shows Uranus undergoing capture into a spin-orbit resonance when Neptune crosses $\sim$24 au en route to its current location at 30 au. Here we have set Neptune's migration rate to $\sim0.05$ au Myr$^{-1}$, which is within the adiabatic limit -- the fastest possible rate to generate a capture with ${\epsilon}_{i}\approx0$\textdegree. The adiabatic limit occurs when Neptune's migration takes it across the resonance width in about a libration time, which is just $2\pi/w_{\textsubscript{lib}}$ with $w_{\textsubscript{lib}} = \sqrt{-\alpha{g}\sin{\epsilon}\sin{I}}$ \citep{2004AJ....128.2510H}. Just as slow changes to the support of a swinging pendulum do not alter the pendulum's motion, gradual changes to Neptune's orbit do not change the behavior of the libration. However, if Neptune's migration speed exceeds the adiabatic limit, then the resonance cannot be established. The top panel of Figure \ref{capture} shows Uranus tilting to 60\textdegree~in 150 Myr when Neptune reaches its current location, and all the way to 90\textdegree~in 600 Myr if we allow Neptune to continue outwards. Planets migrate by scattering planetesimals, which can decrease inclinations; accordingly, we optimistically assumed an initial value for Neptune's inclination at twice its current value. Because we have increased Neptune's inclination and moved Neptune out as fast as possible and yet still allowed capture, 150 million years represents a rough lower limit to the time needed to tilt Uranus substantially. 

The bottom panel of Figure \ref{capture} and Figure \ref{polar:capture} both show the evolution of the resonance angle, and the angle oscillates with a libration period of about 30 Myr about the equilibrium point. The libration period increases as $\epsilon$ increases in accordance with the predictions of $w_{\textsubscript{lib}}$. The noticeable offset of the equilibrium below $\Psi=0$\textdegree~in Figures \ref{capture} and \ref{polar:capture} is due to the rapid migration of Neptune \citep{2004AJ....128.2510H}:

\begin{equation} \label{eqangle}
\Psi_{eq} = \frac{\dot{\alpha}\cos\epsilon\,+\,\dot{g}\cos I }{\alpha g\sin\epsilon\sin I}.
\end{equation}
Recall that $g$, the nodal precession frequency, is negative, $\alpha$ is positive, and as Neptune migrates away from the Sun, $\dot{g}$ is positive. Because $\alpha$ is constant, $\dot{\alpha}=0$, and so $\Psi_{eq}$ is slightly negative in agreement with Figure \ref{capture}. We conclude that although a spin-orbit resonance with Neptune can tilt Uranus over, the model requires that Uranus be pinned between Jupiter and Saturn for a few hundred million years, yet close encounters with either gas giant makes it unstable to leave Uranus there for more than a million years \citep{1973Icar...20..422L,1989Icar...79..223F,1990AJ....100.1680G,1993AJ....105.1987H}. Is there any room for improvement?

Solar system evolution models calculate planetary migration timescales on the order of $10^{6}-10^{7}$ yr \citep{1999AJ....117.3041H, 1999Natur.402..635T,2002AJ....123.2862T,2003Icar..161..431T, 2005Natur.435..466G, 2005Natur.435..462M, 2005AJ....130.2392H, 2005Natur.435..459T, 2020Icar..33913605D}. This is incompatible with this resonance capture scenario, which requires at least $10^{8}$ yr. Speeding up the tilting timescale significantly would require a stronger resonance. The strength of this resonance is proportional to the migrating planet's inclination and it sets the maximum speed at which a capture can occur \citep{1994Icar..109..221H}. Although Neptune's initial orbital inclination angle is unknown, a dramatic reduction in the tilting timescale is implausible.

Another possibility is that the gas giants were once closer to the Sun where tidal forces are stronger. Some evidence for this comes from the fact that the giant planets probably formed closer to the snow line \citep{2006Icar..181..178C}, where volatiles were cold enough to condense into solid particles. Shrinking the planets' semi-major axes by a factor of 10\% decreases the resonance location by about 3 au and reduces the obliquity evolution timescale by about 15\%. Although this is an improvement, a timescale on the order of $10^{8}$ yr seems to be the fundamental limit on the speed at which a significant obliquity can be reached \citep{2016DPS....4831809R,2018CeMDA.130...11Q}. 

Less critical than the timescale problem but still important is the inability of the obliquity to exceed 90\textdegree$\,$ (Figure \ref{capture}). The reason for this follows from Equation \ref{period}, which shows that Uranus's precession period approaches infinity as $\epsilon$ approaches 90\textdegree. Neptune's migration speed then is faster than the libration timescale and the resonance ceases. This effect is more apparent in Figure \ref{polar:capture} in which the libration period also increases with the obliquity. The resonance breaks when the resonance angle stops librating about an equilibrium point and instead circulates a full $2\pi$ radians. \cite{2018CeMDA.130...11Q} show that a related resonance that occurs when the planets are also close to a mean-motion resonance could tilt the planet past 90\textdegree, but this, like the resonance considered here, is probably too weak. Keeping Uranus between Jupiter and Saturn for $10^{8}$ yr is as implausible as the planet having once had a massive distant moon \citep{2010ApJ...712L..44B}.

\section{Obliquity Kick from a Secular Spin--Orbit Resonance} \label{rk}

\begin{figure}[h]
	\centering
	\includegraphics[width=0.7\textwidth]{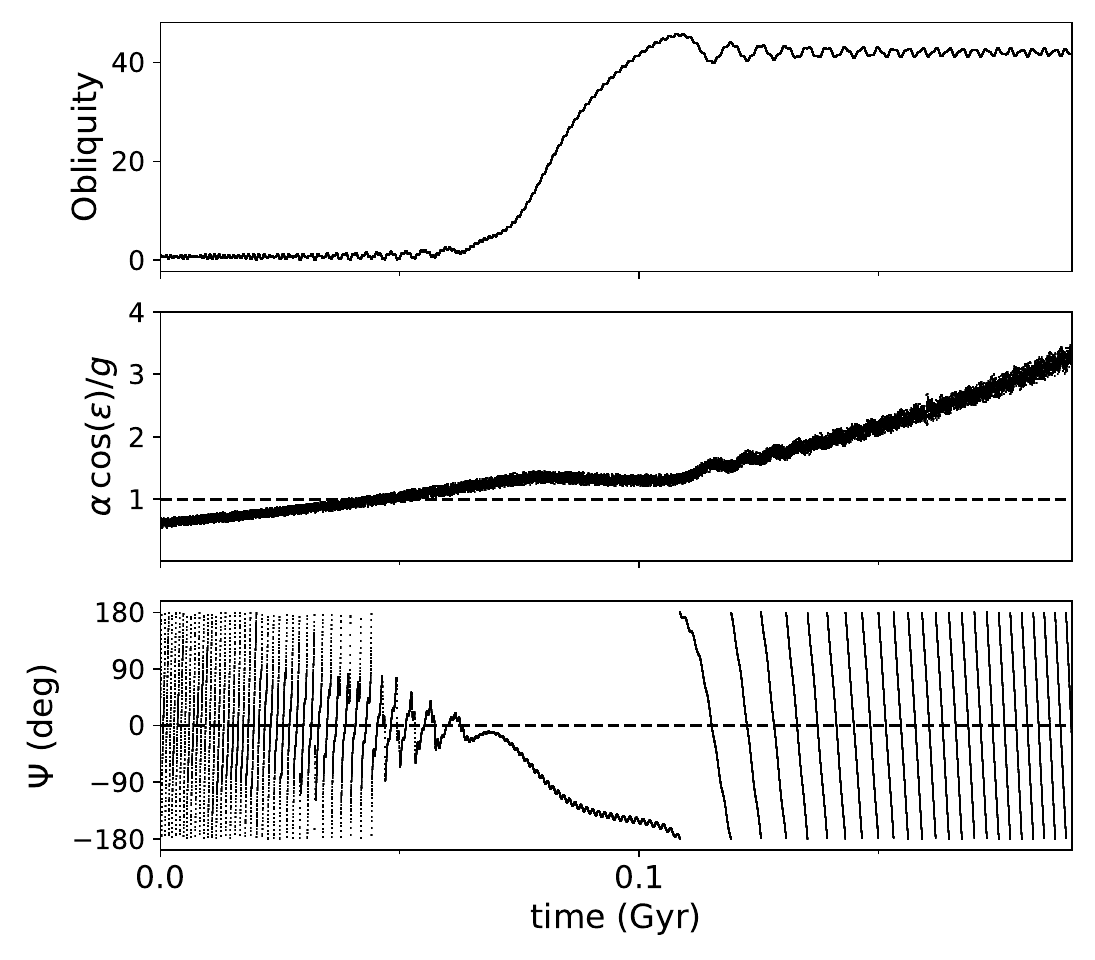}
	\caption{A resonance kick with a particularly large 40\textdegree$\,$ amplitude. Here, Neptune is migrating out at an average speed of 0.068 au Myr$^{-1}$, and Uranus's radius is at its current size. Jupiter, Saturn, and Uranus are located 10\% closer to the Sun than today, and Neptune has an inclination of 4\textdegree.}
	\label{kick}
\end{figure}

A resonance capture with Neptune may not be able to tilt Uranus effectively, but this resonance may still contribute significantly on a timescale more compatible with current planetary formation models. A resonance kick occurs if Neptune's migration speed is too fast to permit captures (i.e. exceeds the adiabatic limit). If $\dot{g}$, the rate Neptune's nodal precession frequency changes as the planet migrates, is large enough, then from Equation \ref{resdt}, $g\cos I$ shrinks faster than Uranus's spin precession frequency $\alpha\cos\epsilon$. Thus, $\dot{\Psi} < 0$, which drives $\Psi$ to -180\textdegree. For a capture, on the other hand, $\dot{g}$ is smaller so that the resonance lasts more than one libration cycle. A kick can also occur at slower migration speeds if the relative phase of the two precession axes are misaligned. Figure \ref{kick} shows an example of a resonance kick with a concurrent change in obliquity lasting 50 Myr. Overall, the magnitude of the kick depends on Neptune's orbital inclination, Uranus's initial obliquity, the migration speed, and the relative orientation of Uranus's spin axis and Neptune's orbital pole at the time the resonance is encountered. We will explore the entirety of this phase space to examine how effective Neptune's resonant kicks are at tilting Uranus.

For a range of migration speeds consistent with orbital evolution rates from planetesimal scattering \citep{1999AJ....117.3041H,2005AJ....130.2392H}, we ran simulations for initial obliquities ranging from $\epsilon\approx0$\textdegree~to $\epsilon\approx90$\textdegree~in increments of 5\textdegree. While Uranus may have originated with zero obliquity due to gas accretion, this does not need to be the case in general. Impacts, for example, are a source of at least small obliquities, the prior spin-orbit resonance discussed by \cite{2020ApJ...888...60R} likely induced significant obliquity, and tidal torques onto detached circumplanetary disks may also excite planetary obliquities \citep{2020ApJ...898L..26M}. For each initial obliquity, we sample a range of phase angles from 0 to $2\pi$. 

\begin{figure}[h]
	\centering
	\includegraphics[width=0.7\textwidth]{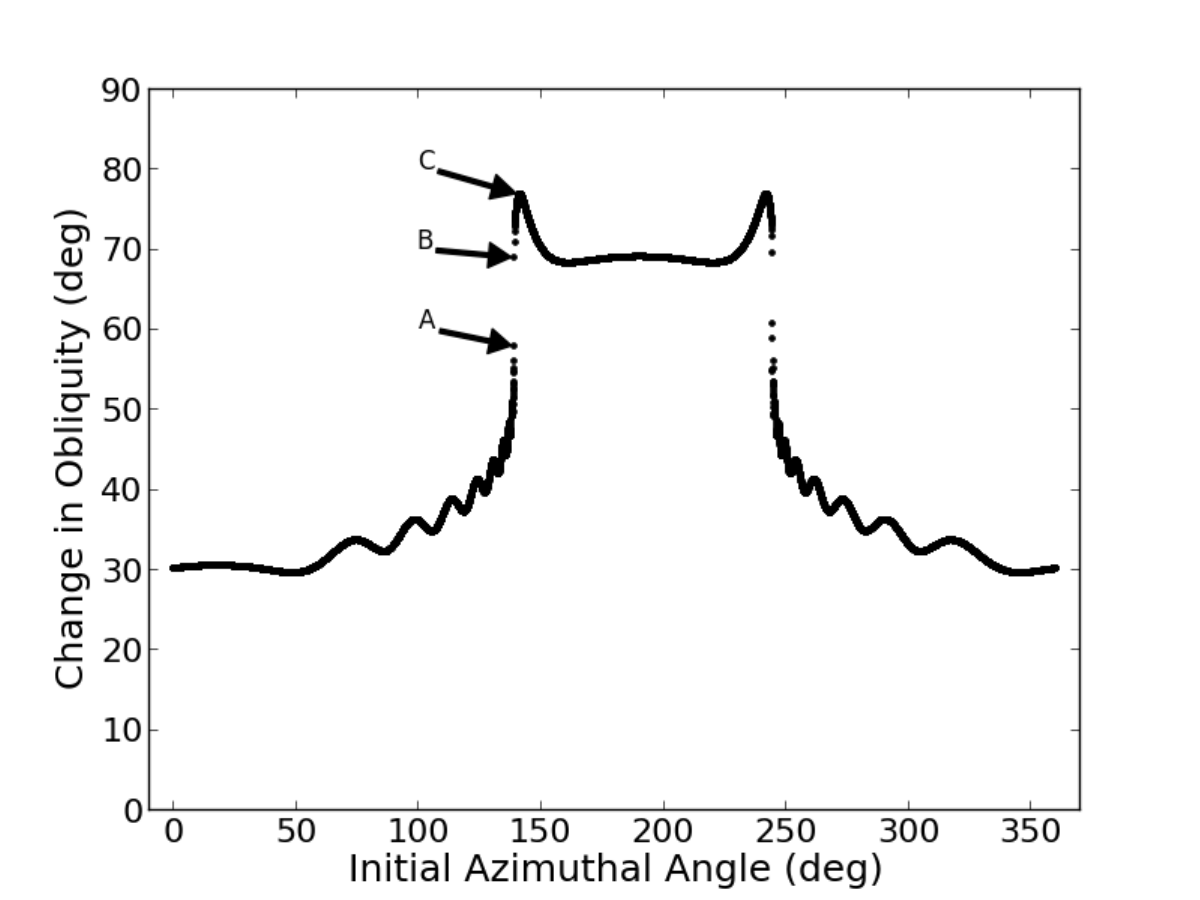}
	\caption{Change in obliquity as a function of Uranus's initial azimuthal angle where $\epsilon=\,$1\textdegree, $i_{N}=\,$8\textdegree, and the system is near the adiabatic limit. Here, we sampled 10,000 initial azimuthal angles from 0\textdegree$\,$ to 360\textdegree$\,$ and raised Neptune's inclination even further to emphasize the transition region from kicks (phases near 0\textdegree) to captures (phases near 180\textdegree). The annotated points (A, B, C) are discussed further in Figure \ref{polar}.}
	\label{phase}
\end{figure}

\begin{figure}[h]
	\centering
	\begin{tabular}[b]{@{}p{0.45\textwidth}@{}}
		\centering
		\includegraphics[width=0.5\textwidth]{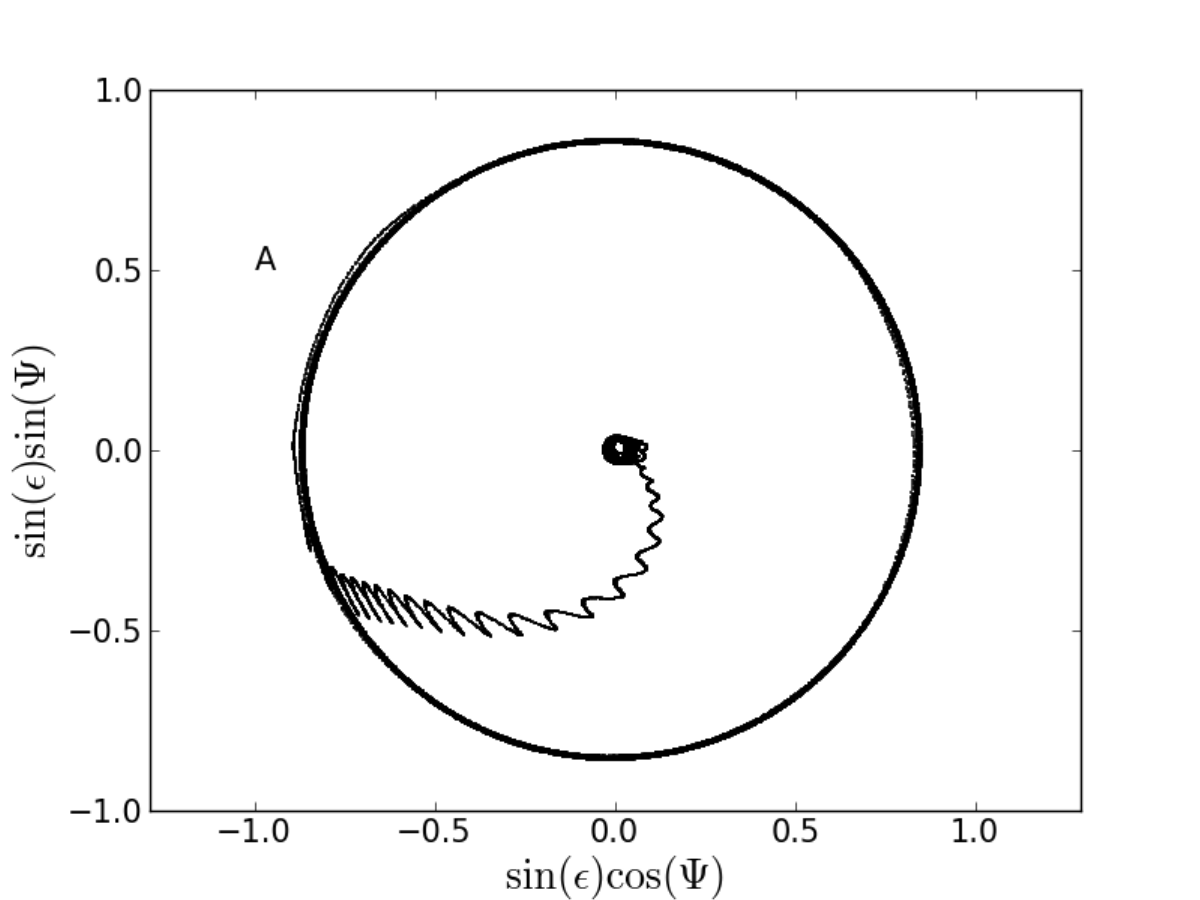}
		\centering (a)
	\end{tabular}\\
	\begin{tabular}[b]{@{}p{0.45\textwidth}@{}}
		\centering
		\includegraphics[width=0.5\textwidth]{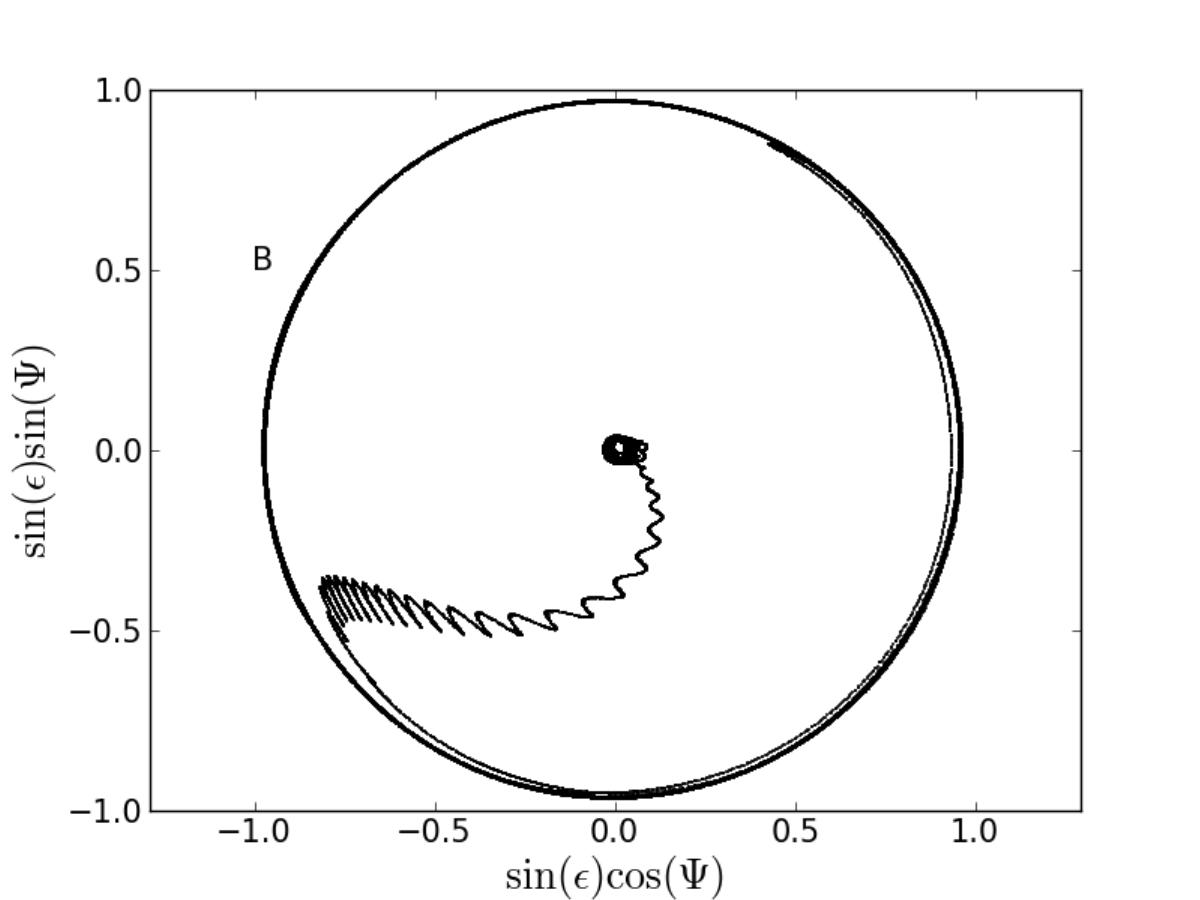}
		\centering (b)
	\end{tabular}\\
	\begin{tabular}[b]{@{}p{0.45\textwidth}@{}}
		\centering
		\includegraphics[width=0.5\textwidth]{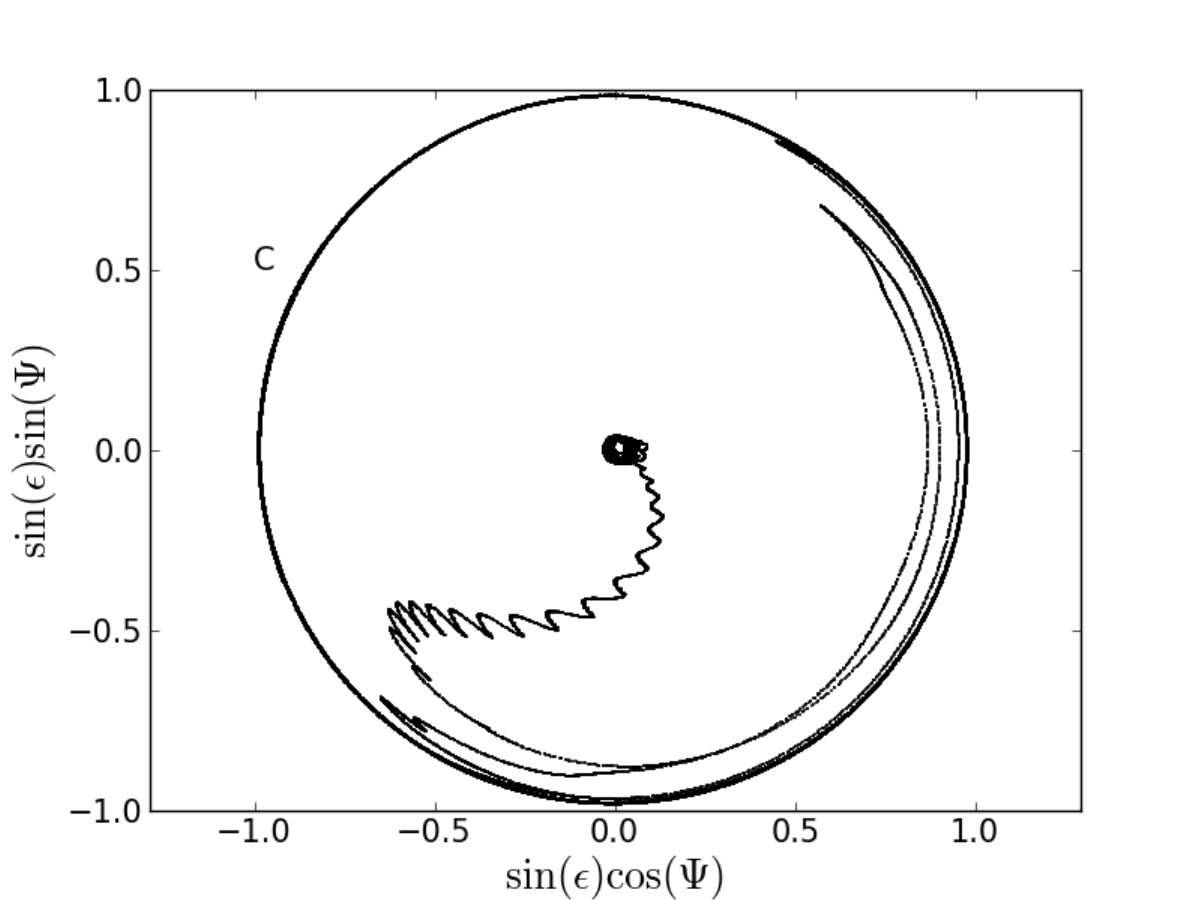}
		\centering (c)
	\end{tabular}

	\caption{These are polar plots of one kick (a) and two captures (b, c) taken from Figure \ref{phase}. (A) The largest resonance kick at the transition region in Figure \ref{phase}. The resonance angle undergoes less than one libration cycle. It approaches 180\textdegree$\,$ and then leaves the resonance. (B) A very tenuous capture whose libration angle exceeds 180\textdegree~for one cycle before escaping the resonance creating the large outer circle. (C) A resonance capture well within the capture region in Figure \ref{phase}. Here the system also breaks free from the resonance after a few libration cycles. Short-period oscillations in these plots are due to the effects of pole precession.}
	\label{polar}
\end{figure}

Distinguishing kicks from captures is more difficult when Neptune is migrating near the adiabatic limit, especially at low inclinations, so to highlight this effect, we raise Neptune's inclination to 8\textdegree~in Figure \ref{phase}. This figure shows how the phase angle determines whether the resonance would yield a kick or a capture. Note, however, that it is actually the phase angle on encountering the resonance that matters, not the initial phase angle plotted in Figure \ref{phase}. Also, the outlying oscillations in this figure are due to librational motion as the final obliquity is calculated only when Neptune reaches its current location at 30 au. In this case, there is a clear division between captures and kicks near azimuthal angles 150\textdegree~and 250\textdegree. In other cases at lower inclinations, however, the boundaries between kicks and captures seem more ambiguous.

Figure \ref{polar} shows the corresponding polar plots for a selection of points in Figure \ref{phase} contrasting the difference between kicks and captures. Near the adiabatic limit, the phase angle will not librate more than one or two cycles for captures before the resonance breaks. This is most apparent in Figure \ref{polar}(b), where Uranus completes just over one libration cycle. For comparison, Figure \ref{polar:capture} shows a capture well within the adiabatic limit, and here, the phase angle clearly librates multiple times until the planet's obliquity reaches $\epsilon\sim90$\textdegree. We therefore identify kicks as a resonance active for less than one libration cycle. Resonance kicks near the adiabatic limit can also generate large final obliquities, so we will focus our attention to this region in phase space. As shown in Figures \ref{kick} and \ref{phase}, it is possible to generate kicks up to $\Delta\epsilon\sim40$\textdegree~for $i_{N}=4$\textdegree~and $\Delta\epsilon\sim55$\textdegree~for $i_{N}=8$\textdegree.

\begin{figure}[h]
	\centering
	\includegraphics[width=0.7\textwidth]{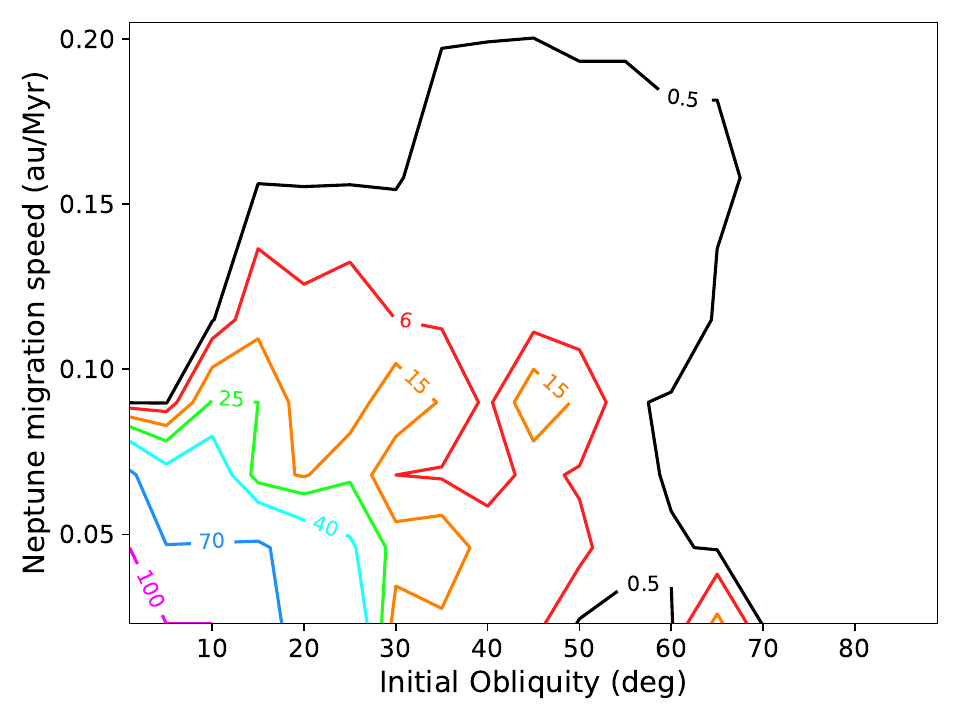}
	\caption{Percentage of resonances that produce captures for a range of initial obliquities and migration speeds. Captures occur most readily in the lower-left corner of the figure for small obliquities and slow migration rates. Here, $i_{N}=4$\textdegree.}
	\label{percentkick}
\end{figure}

In Figure \ref{percentkick}, we map the fraction of resonances that produce captures for a range of migration speeds and initial obliquities. The transition from 100\% kicks to 100\% captures over migration speeds is sharpest at lower initial obliquities. This can be understood by considering the circle that Uranus's spin axis traces as it precesses; for small obliquities, significant misalignments between the two poles are rare, and the outcome of a resonance is determined primarily by Neptune's migration speed. With increasing initial obliquities, large misalignments become more common, and the probability of generating a resonance kick increases \citep{2018CeMDA.130...11Q}.

\begin{figure}[h]
	\centering
	\includegraphics[width=0.7\textwidth]{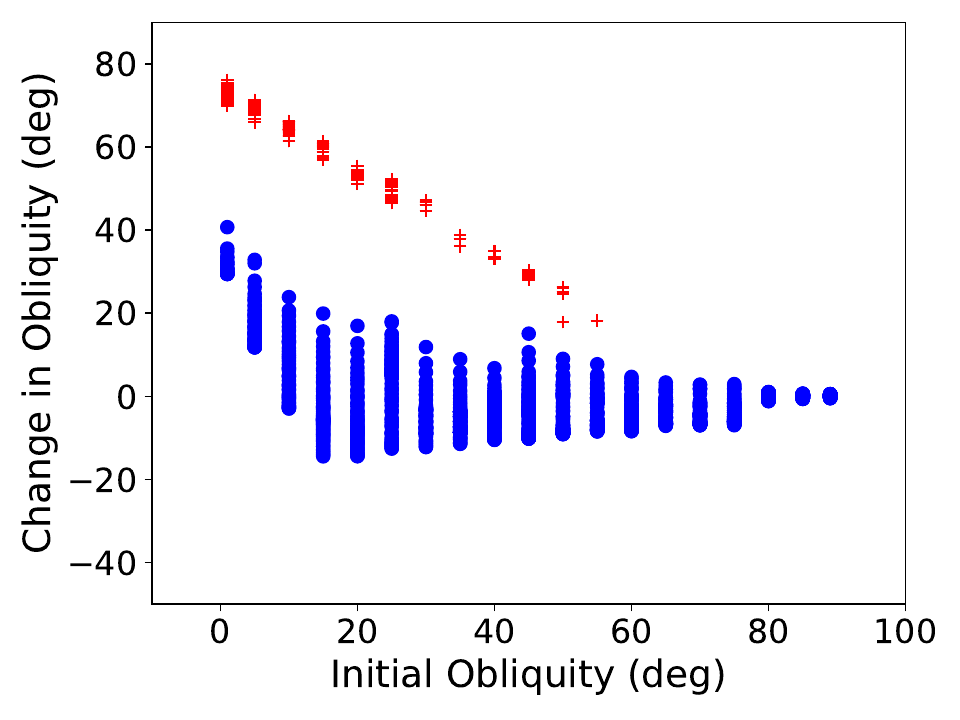}
	\caption{Change in obliquity as a function of Uranus's initial obliquity. The blue circles depict resonance kicks, while the red crosses depict resonance captures. Neptune's migration speed is 0.068 au Myr$^{-1}$, which is near the adiabatic limit at small initial obliquities. We set $i_{N}=4$\textdegree. It should be noted that our sampling of 100 initial azimuthal angles for Uranus is too coarse to resolve any captures for initial obliquities greater than 55\textdegree. It is possible for captures to happen at larger initial obliquities, but the range of favorable phase angles is very small.}
	\label{kickdist}
\end{figure}

We expect and find that the strongest resonant kick occurs at around the adiabatic limit because a slow migration speed gives ample time for the resonance to respond. Conversely, a rapid migration speed would quickly punch through the resonance, leaving little time for the resonance to influence Uranus. Figure \ref{kickdist} depicts the distributions of kicks and captures near the $\epsilon\approx0$\textdegree~adiabatic limit. Looking at the resonance kicks, we see that they can reach maximum changes in obliquities of 40\textdegree~(Figure \ref{kick}) for $i_{N}$ near twice Neptune's current inclination and even greater changes in obliquity for a higher assumed $i_{N}$ (Figure \ref{phase}). This looks promising, but we need to understand the probability of these large kicks. In fact, looking at Figure \ref{kickdist} shows that, for high obliquities, negative kicks are common. For low obliquities, kicks must be positive because $\epsilon$ itself cannot be negative. However, if Neptune is migrating quickly and $\epsilon$ is large enough, then the relative phase angle is random, resulting in a range of possible obliquity kicks; in particular, if $\sin(\Psi)$ is positive in Equation \ref{obldt}, then $\dot{\epsilon}$ is negative. 

\begin{figure}[h]
	\centering
	\begin{tabular}[b]{@{}p{0.45\textwidth}@{}}
		\centering
		\includegraphics[width=0.5\textwidth]{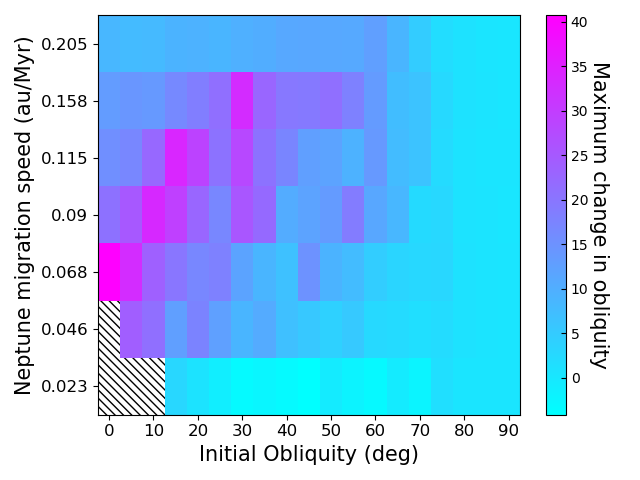}
		\centering (a)
	\end{tabular}\\
	\begin{tabular}[b]{@{}p{0.45\textwidth}@{}}
		\centering
		\includegraphics[width=0.5\textwidth]{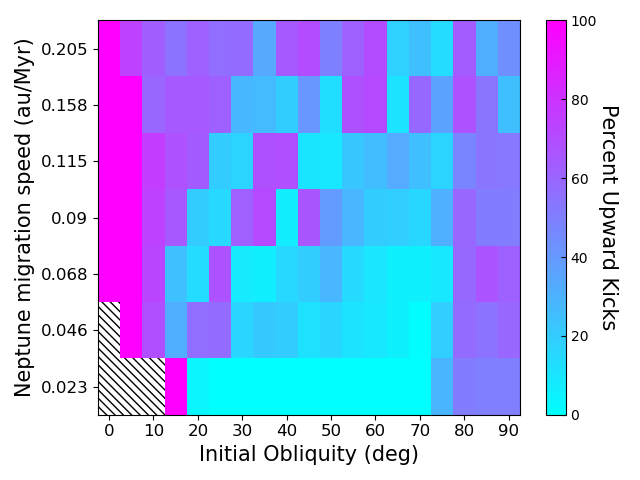}
		\centering (b)
	\end{tabular}\\
	\begin{tabular}[b]{@{}p{0.45\textwidth}@{}}
		\centering
		\includegraphics[width=0.5\textwidth]{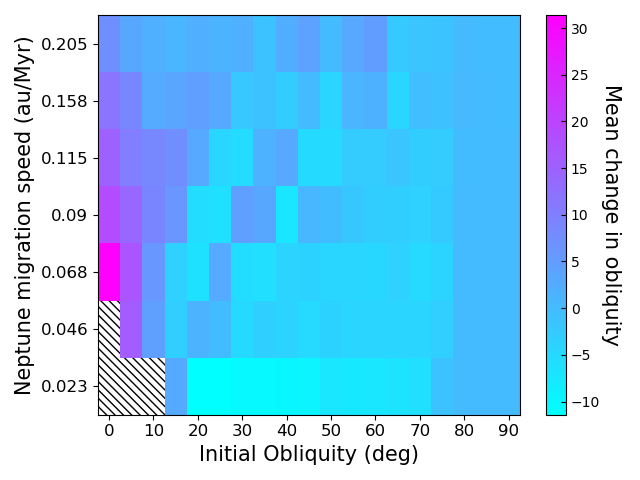}
		\centering (c)
	\end{tabular}
	
	\caption{(a) Corresponding maximum change in obliquity for resonant kicks depicted in Figure \ref{percentkick}. Diagonal hatching in the four boxes to the lower left in all panels corresponds to captures. The scale ranges from 40\textdegree$\,$ (magenta) to 0\textdegree$\,$ (cyan) kicks. (b) Percentage of kicks that yield positive changes in obliquity. 100\% positive kicks are depicted in magenta. (c) Mean changes in obliquity for resonant kicks. The scale measures the change in obliquity, with magenta being the maximum.}
	\label{fig:kickdists}
\end{figure}

\begin{figure}
	\includegraphics[width=0.7\textwidth]{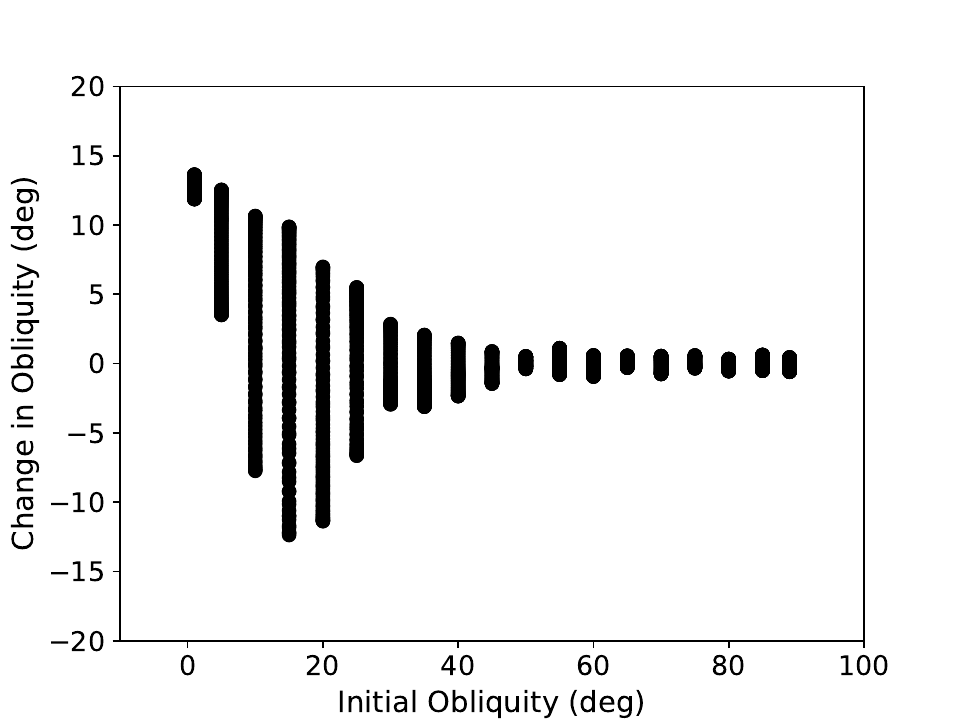}
	\caption{The change in obliquity as a function of Uranus's initial obliquity for a cooling and shrinking Uranus with $i_{N}=4$\textdegree. There are 1900 simulations depicted here.}
	\label{fig:kickdists:d}
\end{figure}

Figure \ref{fig:kickdists}(a) shows the maximum possible kicks over all initial obliquities and migration speeds, and although large kicks are possible, they are rare. Apart from resonant kicks that occur near the adiabatic limit, which can be seen in this figure as the magenta feature extending linearly up and to the right, the maximum strength of resonant kicks is typically $\Delta\epsilon\approx10$\textdegree$\,-\,20$\textdegree. On top of that, resonance kicks can also decrease obliquities, which are depicted in Figure \ref{fig:kickdists}(b). If Uranus's obliquity was initially large, then the percentage of positive kicks is around 50\%, tending toward primarily negative kicks as Neptune's migration speed decreases. Because about half of all possible resonance kicks at initial obliquities greater than 10\textdegree~are negative, the average kick should be low. Figure \ref{fig:kickdists}c depicts the corresponding mean changes in obliquity, and they tend to be weak, with mean resonance kicks of only a few degrees. At low initial obliquities, though, kicks tend to increase the planet's obliquity by at least 10\textdegree. Generating a large resonance kick would most commonly occur if ${\epsilon}_{i}=$ 0\textdegree~with Neptune migrating no faster than 0.1 au Myr$^{-1}$ or a crossing time on the order of $\sim10^{7}$ yr. These figures show that, as a statistical process, resonances have only a weak effect, and that one needs favorable initial conditions for large kicks.

We could increase Uranus's obliquity further if it received multiple successive resonance kicks. This might be achieved, for example, with either a resonance between Uranus and another possible ice giant that may have existed in the \cite{1999Natur.402..635T} model, a resonance with its own orbital pole after Uranus' spin precession rate was amplified by harboring a massive extended circumplanetary disk \citep{2020ApJ...888...60R}, or if Uranus's precession frequency quickened as the planet cools and shrinks. The latter process is interesting and merits further discussion. 

Uranus was hotter and therefore larger in the past \citep{1986Icar...67..391B,1991uran.book..469P,1996Icar..124...62P,2009Icar..199..338L}, and conserving angular momentum requires that a larger Uranus must spin significantly more slowly. Both Uranus's spin angular frequency, $\omega$, and its quadrupole gravitational harmonic, $J_{2}$, appear in Equation \ref{prec} and change if the planet's radius changes. Because $\omega\propto R^{-2}$ and $J_{2}\propto{\omega}^{2}$ \citep{2009ApJ...698.1778R}, the result is a slower precession frequency. Here, for simplicity, we have ignored the contributions of the satellites, as including them would soften the response somewhat. Although this is highly dependent on Uranus's cooling rate, \cite{1986Icar...67..391B} and \cite{1991uran.book..469P} show that Uranus shrank by a factor of 2 on a timescale of order 10 Myr. We simulated this scenario by having Uranus's radius decrease according to an exponential function with Neptune stationary at 25 au. Figure \ref{fig:kickdists:d} shows the resulting kicks as a function of Uranus's initial obliquity, and they never exceed 15\textdegree. Scenarios that include multiple crossings of the same resonance would likely still fall short of fully tilting Uranus \citep[e.g.][]{2004Natur.429..848C,2004AJ....128.2510H,2004AJ....128.2501W}.

\section{Revisiting the Collision Model}

\subsection{Conditions for Collisions}

Recall that the leading hypothesis for Uranus's tilt is a single Earth-mass impactor striking the planet's polar region \citep[e.g.][]{1992Icar...99..167S,2020NatAs...4..880I}, but that \cite{2012Icar..219..737M} argue for two or more collisions if the satellites were formed primordially from a circumplanetary disk \citep{2018ApJ...868L..13S}. In this section, we consider each of these scenarios and derive the resulting probability distributions for such impacts. To do this, we designed a collisional code that builds up a planet by summing the angular momenta of impactors to determine the planet's final obliquity and spin rate under various circumstances, and we typically run this for half a million randomized instances. Our assumptions are that the impactors originate within the protoplanetary disk, they approach a random location on the planet on trajectories that parallel its orbital plane, and all the mass is absorbed upon impact. Because nearly every object in the solar system orbits in roughly the same direction, the impactors' relative speed would be at most several tens of percent of Uranus's orbital speed (6.8 km s$^{-1}$). Because we expect most impactors to follow orbits with lower eccentricities, we sample relative velocities between 0 and 0.4 times Uranus's circular speed.

Considering that the impactor's relative speed is small compared to the planet's escape speed (21.4 km s$^{-1}$), gravitational focusing is also important. For cases where gravitational focusing is strong, the impact cross section is large and the impactor is focused to a hyperbolic trajectory aimed more closely toward the planet's center. Because head-on collisions do not impart any angular momentum, we expect the planet's spin state to be more difficult to change when focusing is included. The impact parameter for this effect is given by $b$ with
\begin{equation}\label{eq:GF}
b^{2}=R_{p}^{2}(1+(V_{\textsubscript{esc}}/ V_{\textsubscript{rel}})^{2}).
\end{equation}

Also, because we do not know how the density profile changes between impacts, we maintain the dimensionless moment of inertia at $K\equiv\frac{I}{M_{p}R_{p}^{2}}=0.225$ but vary the planet's radius as the cube root of the total mass. Although these assumptions are mildly inconsistent, we find that even large impacts incident on a mostly formed Uranus yield just small changes in radius and that the final spin rates changes by only about 10\% for other mass-radius relations. Finally, \cite{2012ApJ...759L..32P} suggest a maximum impact boundary of around 0.95 $R_{p}$ as beyond this, the impactor simply grazes the planet's atmosphere and departs almost unaffected. For simplicity, and in the spirit of approximation, we ignore this subtlety. 

\begin{figure}[h]

	\begin{tabular}[b]{@{}p{0.45\textwidth}@{}}

		\includegraphics[width=0.5\textwidth]{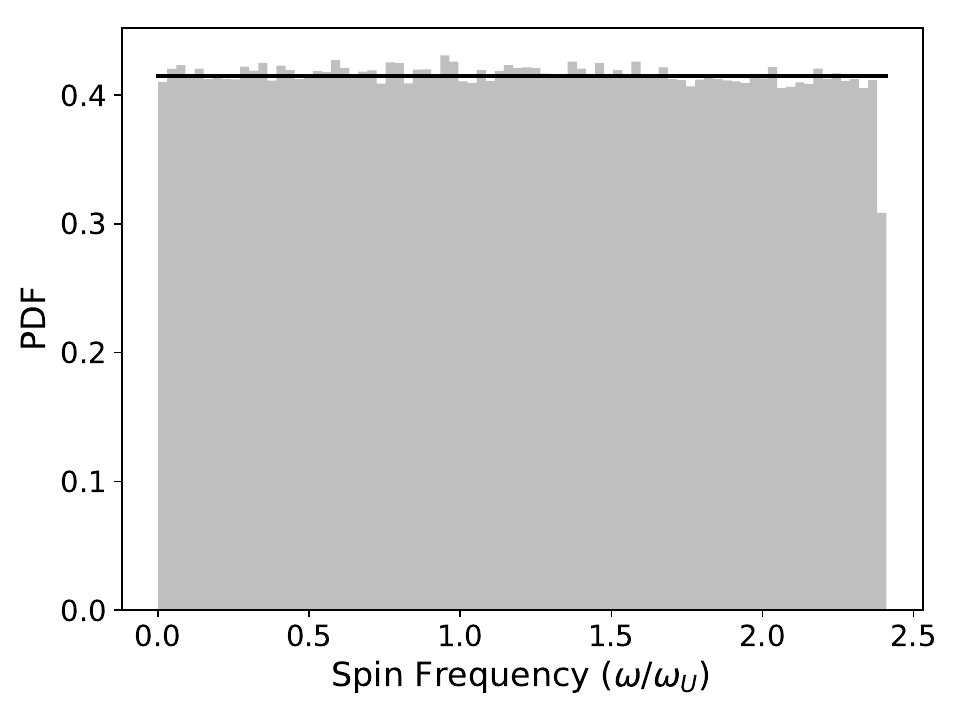}
		\centering (a)
	\end{tabular}
	\begin{tabular}[b]{@{}p{0.45\textwidth}@{}}

		\includegraphics[width=0.5\textwidth]{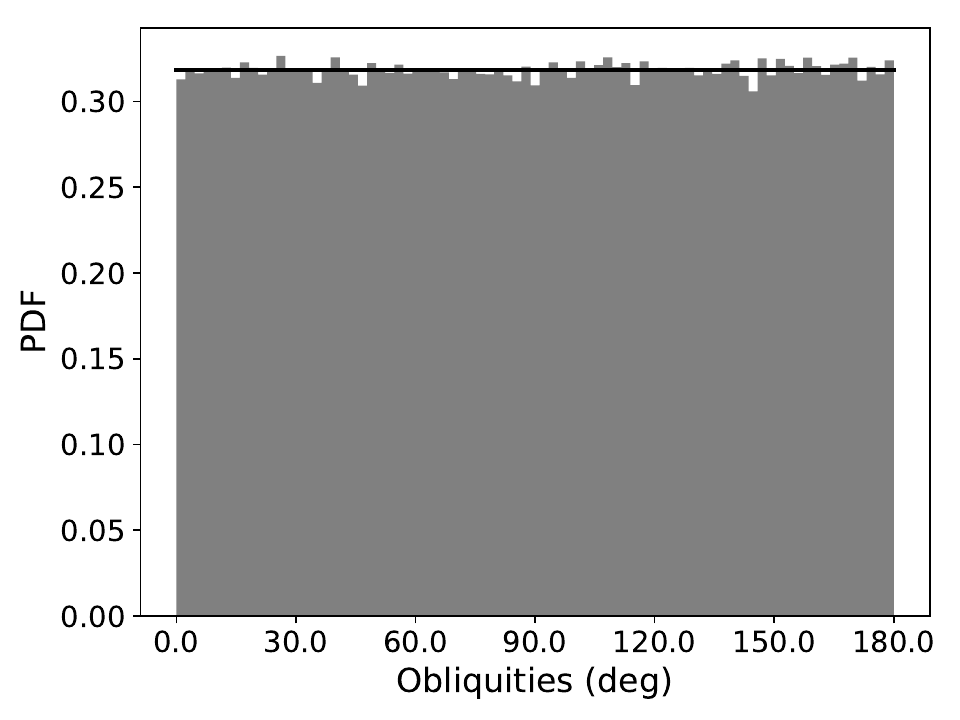}
		\centering (b)
	\end{tabular}
	\begin{tabular}[b]{@{}p{0.45\textwidth}@{}}
		
		\includegraphics[width=0.5\textwidth]{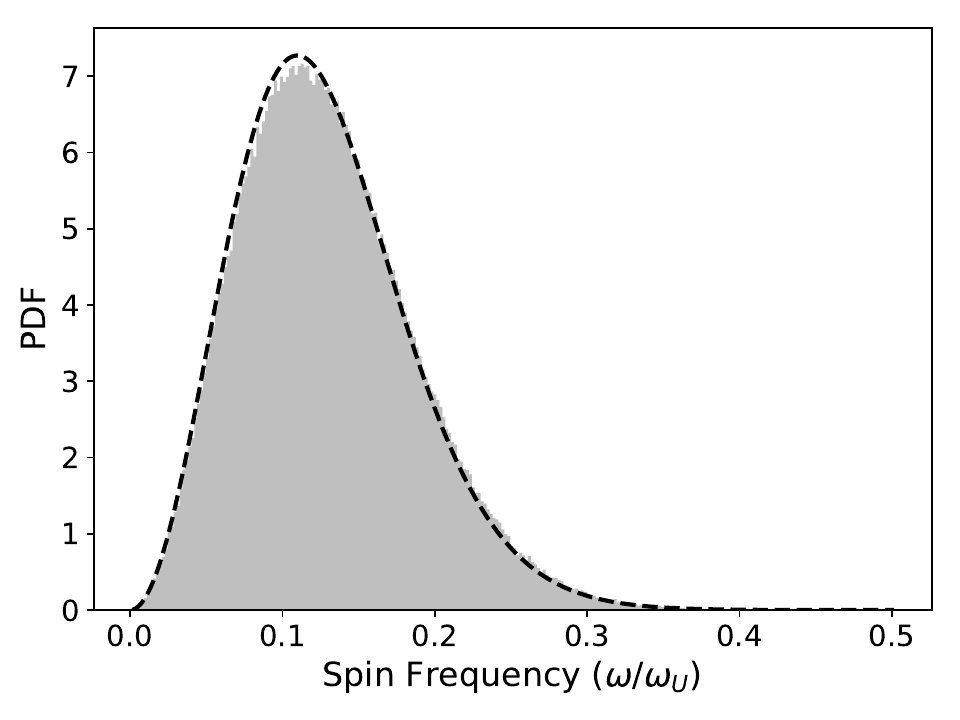}
		\centering (c)
	\end{tabular}
	\begin{tabular}[b]{@{}p{0.45\textwidth}@{}}
		
		\includegraphics[width=0.5\textwidth]{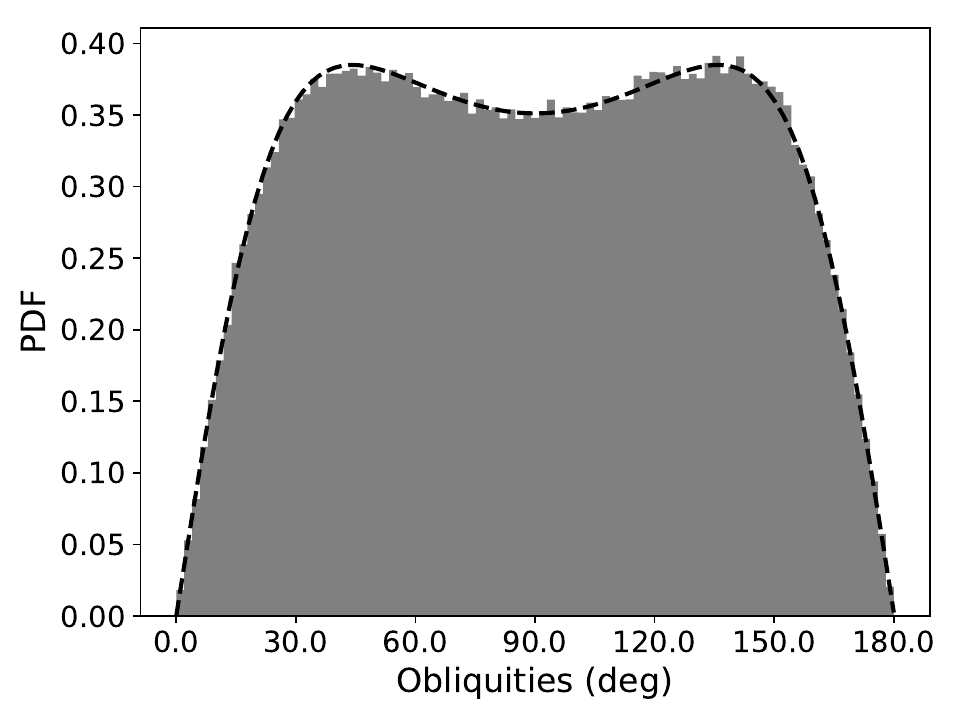}
		\centering (d)
	\end{tabular}

	\caption{(a) The spin distribution for $5\times10^{5}$ realizations of a single impact ($m_{i}=1\,M_{\oplus}$) on a nonspinning proto-Uranus with initial mass $13.5\,M_{\oplus}$ including the effects of gravitational focusing. $\omega_{U}$ is the current uranian spin angular frequency, and all of the following distributions are normalized so that the shaded areas equal 1 (with the obliquities in radians); therefore, the solid line that fits the distribution is the probability distribution function (PDF) $P=\omega_{U}/\omega_{max}$. (b) The corresponding obliquity distribution (depicted in degrees) with the solid line given by $P=1.0/\pi$. (c) The spin distribution for 100 impacts of equal mass ($m_{i}=0.01\,M_{\oplus}$). (d) The corresponding obliquity distribution for 100 impacts. The dashed lines tracing the distributions in both of these figures are the analytic results (Equations (\ref{eq:angmompd:2}) and (\ref{eq:oblpd:1})), and a detailed analysis can be found in the Appendix.}
	\label{fig:onehit}
\end{figure}

\subsection{Accretion of Planetesimals and Protoplanets}

In Figures \ref{fig:onehit}(a) and \ref{fig:onehit}(b), we assume that the planet's initial spin rate was low to highlight the angular momentum imparted by impacts. Because $V_{\textsubscript{esc}}^{2}=2GM_{p}/R_{p}$, the impact cross section $b^{2}\propto R_{p}$ for $V_{\textsubscript{rel}} \ll V_{\textsubscript{esc}}$ (Equation \ref{eq:GF}). The corresponding probability density distribution of impact locations is $\frac{d(\pi b^{2})}{dR_{p}}$, which is constant; therefore, the spin distribution induced from a single collision is flat (Figure \ref{fig:onehit}(a)). However, if the impactor's relative speed is instead much greater than the planet's escape speed, then gravitational focusing is weak, and the impactors will be traveling on nearly straight lines. In this case, a single collision produces a spin distribution that increases linearly, as there is an equal chance of striking anywhere on the planet's surface. But because gravitational focusing only varies the radial concentration of impacts on a planet's surface, the obliquity distribution for a single impact onto an initially nonspinning planet with or without gravitational focusing is uniform. A Uranian core formed from the accretion of many small objects, by contrast, would likely have a very low spin rate \citep{1991Icar...94..126L, 1993Icar..103...67D, 1993Sci...259..350D, 1999Icar..142..219A} because each successive strike likely cancels out at least some of the angular momentum imparted from the previous impact (Figures \ref{fig:onehit}(c) and \ref{fig:onehit}(d)). The planet would also have a narrower range of likely obliquities because the phase space available for low tilts is small.

\begin{figure}[h]
	\centering
	\begin{tabular}[b]{@{}p{0.45\textwidth}@{}}
		\centering
		\includegraphics[width=0.45\textwidth]{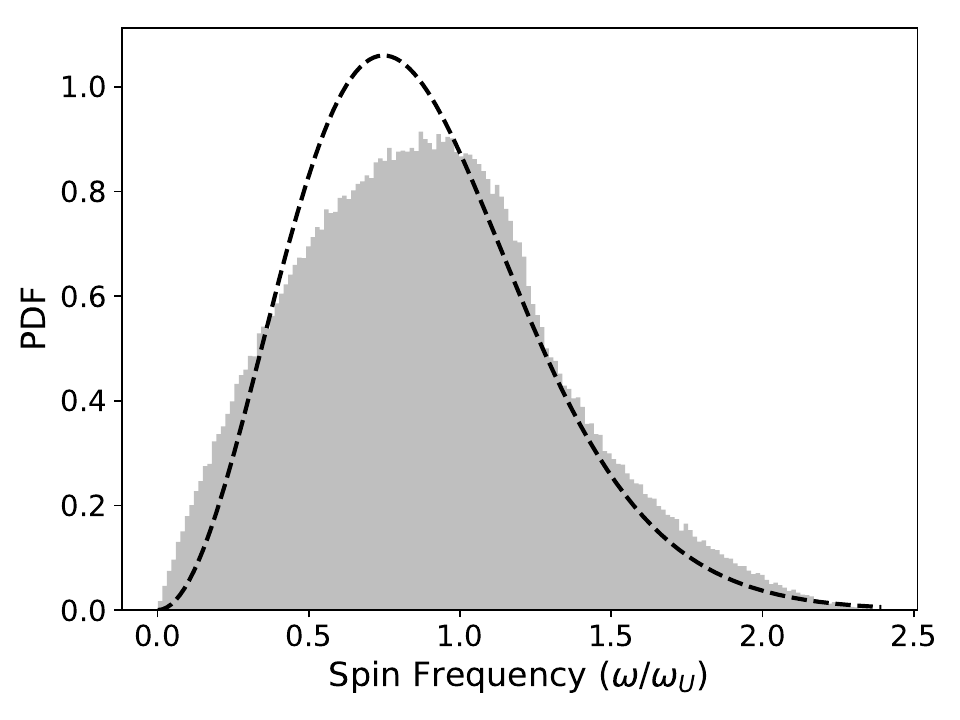}
		\centering (a)
	\end{tabular}\\
	\begin{tabular}[b]{@{}p{0.45\textwidth}@{}}
		\centering
		\includegraphics[width=0.45\textwidth]{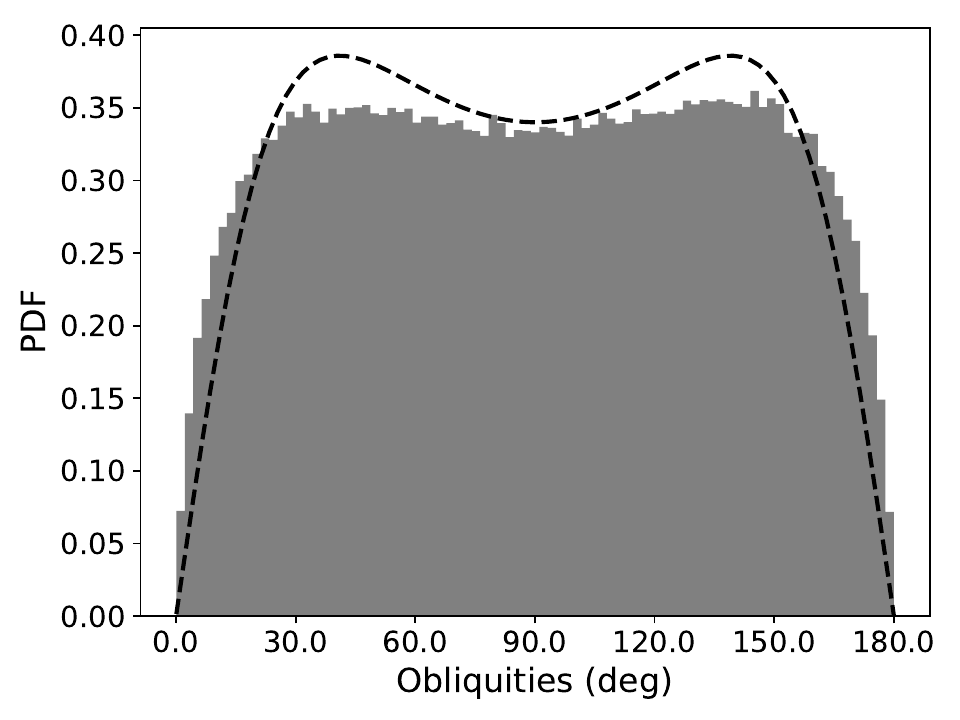}
		\centering (b)
	\end{tabular}\\
	\begin{tabular}[b]{@{}p{0.45\textwidth}@{}}
		\centering
		\includegraphics[width=0.45\textwidth]{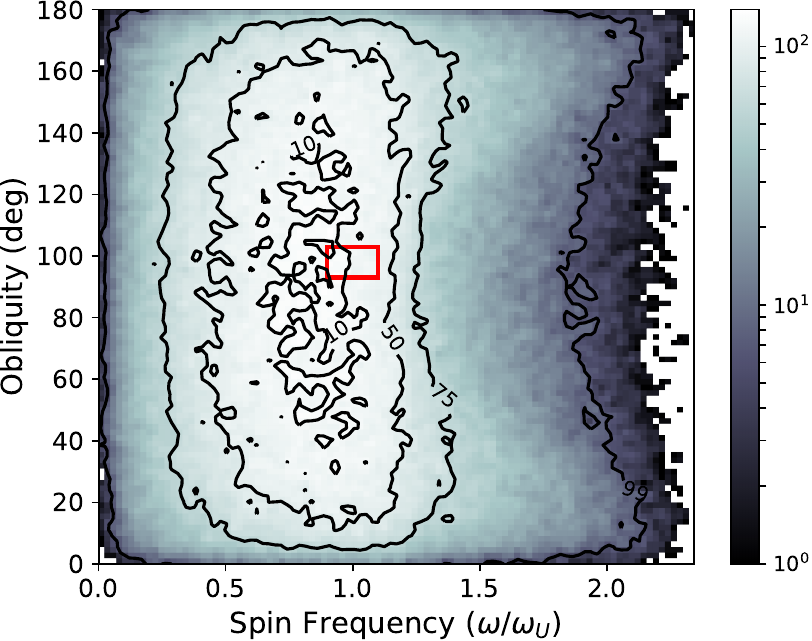}
		\centering (c)
	\end{tabular}
	
	\caption{(a) The spin distribution for two impacts of equal mass ($m_{i}=0.5\,M_{\oplus}$) onto an initially nonspinning Uranus. (b) The corresponding obliquity distribution for two equal impacts. The dashed line is the analytic result for the limit of an Earth mass distributed amongst a large number of particles. (c) A density plot of the spin frequency vs. obliquity where the value of each pixel is the number of iterations that yielded that result. Values within 10\% of Uranus's current obliquity and spin rate are contained within the red rectangle. The probability of falling within this rectangle compared to a similar space around the most likely value is 0.96, meaning that the current state is a likely outcome. }
	\label{fig:twoequal}
\end{figure}

\begin{figure}[h]
	\centering
	\begin{tabular}[b]{@{}p{0.45\textwidth}@{}}
		\centering
		\includegraphics[width=0.45\textwidth]{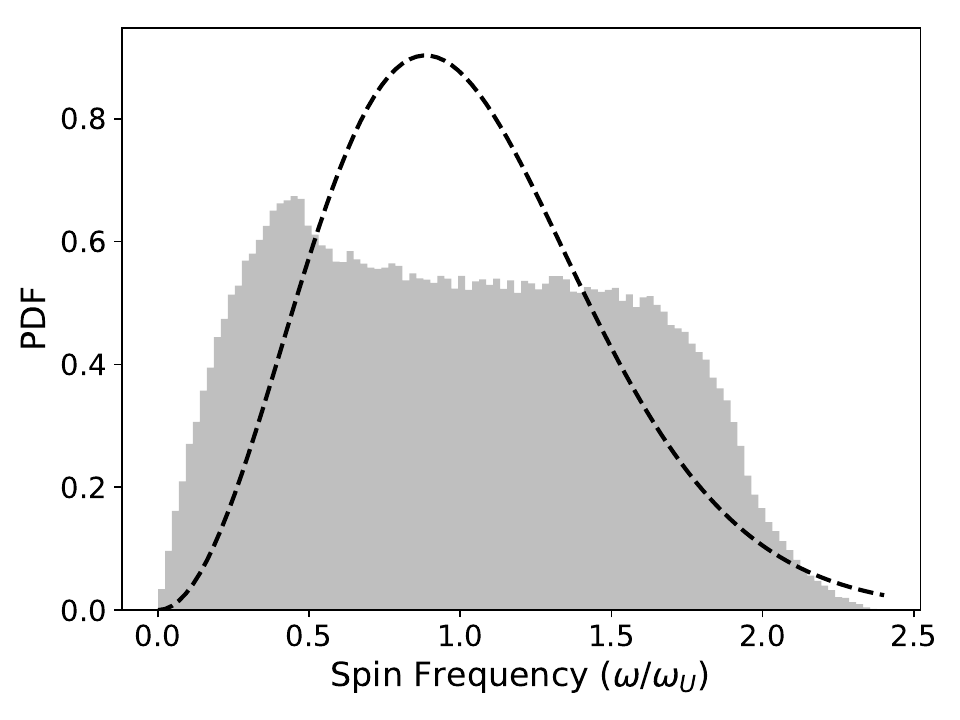}
		\centering (a)
	\end{tabular}\\
	\begin{tabular}[b]{@{}p{0.45\textwidth}@{}}
		\centering
		\includegraphics[width=0.45\textwidth]{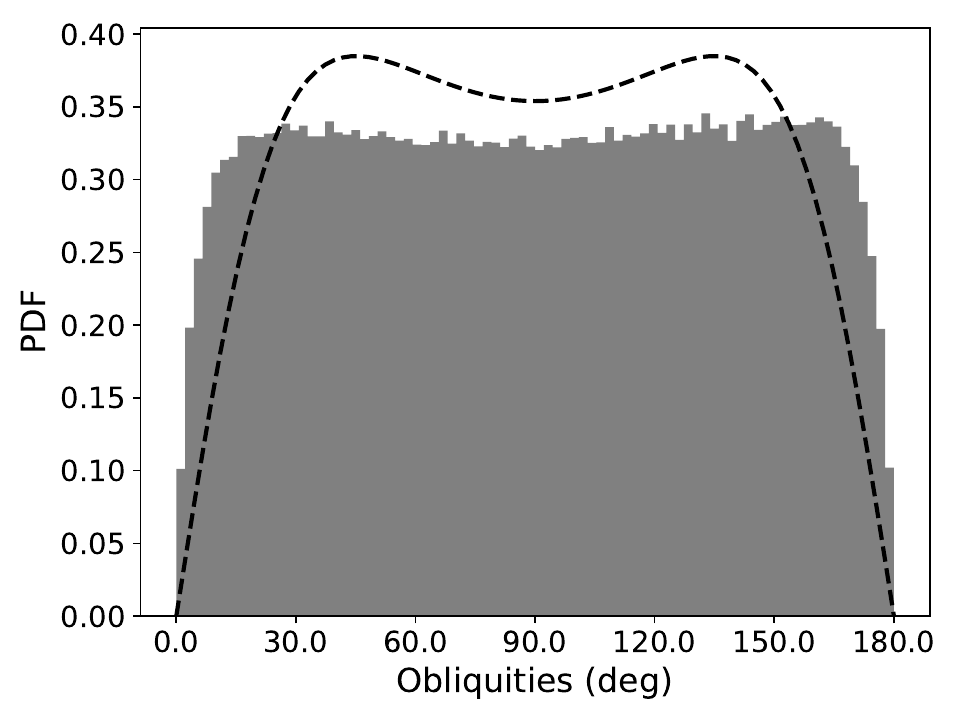}
		\centering (b)
	\end{tabular}\\
	\begin{tabular}[b]{@{}p{0.45\textwidth}@{}}
		\centering
		\includegraphics[width=0.45\textwidth]{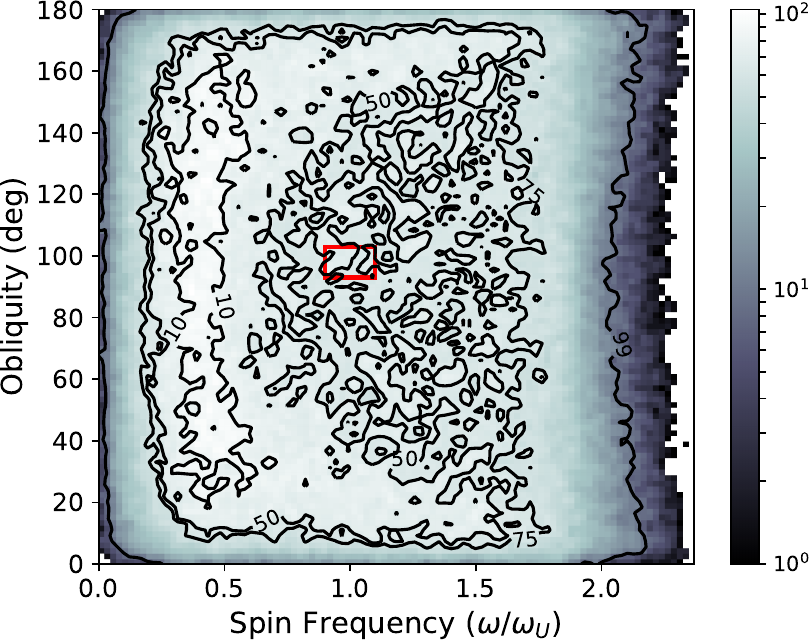}
		\centering (c)
	\end{tabular}
	
	\caption{(a) The spin distribution for two impacts of masses $0.8\,M_{\oplus}$ and $0.2\,M_{\oplus}$ onto a nonspinning planet. (b) The corresponding obliquity distribution for these two unequal impacts. The dashed line is the analytic result for the limit of an Earth mass distributed amongst a large number of particles. (c) A density plot of the spin frequency vs. obliquity where each pixel is the number of iterations that yielded those values. Values within 10\% of Uranus's current obliquity and spin rate are contained within the red rectangle. The likelihood of falling within 10\% of the planet's current spin state is $l_{U}=0.0062$, 0.76 times that of falling within 10\% of the most likely value.}
	\label{fig:twounequal}
\end{figure}

The calculation for the planet's final spin state for many impacts behaves similarly to a random walk, so from the central limit theorem, each directional component of the imparted angular momentum can be described by a normal distribution. The theoretical curve of Figure \ref{fig:onehit}c is given by the probability distribution $f_{{L}}(l)$, which describes the probability that $L$, the magnitude of the planet's spin angular momentum $L = \sqrt{L_{X}^{2} + L_{Y}^{2} + L_{Z}^{2}}$, takes the value $l$:
\begin{equation}\label{eq:angmompd:2}
f_{{L}}(l) = \frac{2l^{2}e^{-l^{2}/2{\sigma}^{2}}}{\sqrt{2{\pi}}\,{\sigma}^{2}{\sigma}_{z}}\,\Phi(0.5;1.5;-\beta l^{2})
\end{equation}
\citep[Equation (109)]{1993Icar..103...67D}. Here, $\sigma$ is the standard deviation for the components of the planet's spin angular momentum that lie in the orbital plane, $\sigma_{z}$ is the standard deviation for the component perpendicular to the orbital plane, and $\beta=\frac{{\sigma}^{2} - {\sigma}_{z}^{2}}{2{\sigma}^{2}{\sigma}_{z}^{2}}$. The angular momentum imparted is always perpendicular to the plane of the impactor's trajectory. After multiple impacts, standard deviations are related by $\sigma_{z} \approx \sqrt{2}\sigma$, so $\beta<0$. Finally, $\Phi(0.5;1.5;\beta l^{2})$ is the confluent hypergeometric function of the first kind. The corresponding obliquity probability distribution is
\begin{equation}\label{eq:oblpd:1}
f_{\epsilon}(\varepsilon)=\left|\frac{1}{4\sqrt{2}\,{\sigma}^{2}{\sigma}_{z}}\frac{\tan(\varepsilon)}{\cos^{2}(\varepsilon)}{{\left(\frac{\tan^{2}(\varepsilon)}{2{\sigma}^{2}} + \frac{1}{2{\sigma}_{z}^{2}}\right)}^{-3/2}}\right|
\end{equation}
\citep[Eq. 111]{1993Icar..103...67D}; we provide derivations of these two equations in the Appendix. Notice how well these calculations agree with the numerical result for many impacts (Figures \ref{fig:onehit}c and \ref{fig:onehit}d). Consequentially, keeping the total mass imparted constant and increasing the number of impactors in Figure \ref{fig:onehit}c from 100 to 1000 would shift the peak to slower spin rates by a factor of $\sqrt{10}$. Because Uranus's spin period is quite fast, its spin state could not have simply been a by-product of myriad small collisions.

Accordingly, we will now consider the intermediary cases with only a few impactors incident on a nonspinning planet. Figure \ref{fig:twoequal} shows the product of two equal-sized hits, and the resulting distributions already resemble the limit of multiple collisions. If the masses of the two impactors differ significantly, however, the corresponding spin and obliquity distributions are more similar to those of the single impact case (Figure \ref{fig:twounequal}). Therefore, while the planet's obliquity distribution may be more or less flat, its spin rate strongly depends on both the number of strikes and the total mass in impactors.

		

\begin{table}[h]
	\begin{tabular}{l|lr|l|c}
		N   & $M_{i}$  & $M_{T}$   & Probability ($l_{U}$) & Normalized Probability\\ \hline
		1   & 1        & 1.0 &  5.0$\times 10^{-3}$ & 1.00\\
		2   & 0.5      & 1.0 &  1.1$\times 10^{-2}$  & 2.20\\
		3   & 0.333    & 1.0 &  7.1$\times 10^{-3}$  & 1.42\\ 
		4	& 0.25	   & 1.0 &  4.5$\times 10^{-3}$  & 0.90\\
		7   & 0.142    & 1.0 &  6.4$\times 10^{-4}$  & 0.13\\
		100 & 0.01     & 1.0 &  0 & 0\\ \hline
		2   & 0.8, 0.2 & 1.0 &  6.2$\times 10^{-3}$  & 1.24\\ \hline
		1   & 0.41      & 0.41 &  5.2$\times 10^{-3}$  & 1.04\\
		2   & 0.205      & 0.41 &  4.4$\times 10^{-5}$  & 0.001\\
		3   & 0.137    & 0.41 &  2.0$\times 10^{-6}$ & $\sim$0\\ \hline
		1   & 3.4      & 3.4 &  1.6$\times 10^{-3}$  & 0.32\\
		2   & 1.7      & 3.4 &  2.3$\times 10^{-3}$ & 0.46
		
	\end{tabular}
	\caption{A Nonrotating Uranus}{\textbf{Note.} This table shows the probability of a number of collisions (N) each with mass $M_{i}$ totaling to $M_{T}$ (in Earth masses) simultaneously generating a spin rate between $0.9<\omega/\omega_{U}<1.1$ and an obliquity between $93^{\circ}<\epsilon<103^{\circ}$ out of $5\times10^{5}$ realizations. In this data set, Uranus is initially non-spinning with an obliquity of 0\textdegree, and in general, probabilities decrease with more impactors. The final column divides the probability by the odds of generating Uranus's current state from a single Earth-mass impactor (first entry).}
	\label{table:odds:1}
\end{table}

Table \ref{table:odds:1} shows a range of possible collisions onto a nonspinning planet. Here we show that the smallest amount of mass necessary to push Uranus toward its observed spin state is about $0.4\,M_{\oplus}$, regardless of the number of impacts. The odds of this happening decrease for each additional collision because each impact needs to hit at exactly the right location. We also provide statistics for impactors much greater than an Earth-mass in the last section of Table \ref{table:odds:1}. Impactors this massive would likely violate our no mass-loss assumption, yet the odds of generating Uranus's current spin state are still low. A more detailed analysis of these impacts is beyond the scope of this paper; however, see \cite{2018ApJ...861...52K,2019MNRAS.487.5029K} for a smooth particle hydrodynamics analysis on the effects that impacts have on Uranus's rotation rate and internal structure. 

We also explored cases with multiple unequal-sized impactors and discovered that the order of the impacts does not matter, as expected, and that the odds are improved for similar-sized impactors. An example of this can be seen in Figures \ref{fig:twoequal}(a) and \ref{fig:twounequal}(a) where for the same total mass, the spin distribution for two equally sized impactors is concentrated near Uranus's current spin state, whereas the distribution is flatter for two unequal-sized impacts. We conclude that a small number of equal impacts totaling to about $1\,M_{\oplus}$ is the most likely explanation for Uranus's spin state if the planet was initially nonspinning.

\subsection{Adding the Effects of Gas Accretion}

Gas accretion almost certainly provides a significant source of angular momentum, so much so that we might expect the giant planets to be spinning at near break-up velocities if they accreted gas from an inviscid thin circumplanetary disk \citep{1986Icar...67..391B, 2009Icar..199..338L, 2010AJ....140.1168W}. Instead, we observe the gas giants to be spinning several times slower, so there must have been some process for removing excess angular momentum. This mechanism may be a combination of multiple effects: magnetic braking caused by the coupling between a magnetized planet and an ionized disk \citep{2011AJ....141...51L, 2018AJ....155..178B}, vertical gas flow into the planet's polar regions and additional midplane outflows from a thick circumplanetary disk \citep{2012ApJ...747...47T,2014ApJ...782...65S}, and magnetically driven outflows \citep{1998ApJ...508..707Q,2003A&A...411..623F,2012ApJ...749L..37L,2013ApJ...779...59G}. Because both Uranus and Neptune spin at about the same rates and have likely harbored circumplanetary disks of their own \citep{2018ApJ...868L..13S}, we suspect that gas accretion is responsible, though pebble accretion may also contribute a significant amount of prograde spin \citep{2020Icar..33513380V}. As such, the planet's initial obliquities should be near 0\textdegree~as the angular momentum imparted by gas accretion is normal to the planet's orbital plane.

\begin{table}[h]
	\begin{tabular}{l|lcr|l|c}
		N   & $M_{i}$       & $M_{T}$  & $\epsilon_{i}$  & Probability ($l_{U}$) & Normalized Probability  \\ \hline
		1	& 1.0      & 1.0  & 0\textdegree  & 4.5$\times 10^{-3}$ &  0.90\\
		2   & 0.2      & 0.5  & 0\textdegree  & 5.4$\times 10^{-4}$ &  0.11\\
		2   & 0.5      & 1.0  & 0\textdegree  & 1.0$\times 10^{-2}$ &  2.00\\
		2   & 1.0      & 2.0  & 0\textdegree  & 4.7$\times 10^{-3}$ &  0.94\\
		2   & 1.5      & 3.0  & 0\textdegree  & 2.5$\times 10^{-3}$ &  0.50\\ \hline
		1	& 1.0      & 1.0  & 40\textdegree & 4.7$\times 10^{-3}$ &  0.94\\
		2   & 0.25     & 0.5  & 40\textdegree & 9.0$\times 10^{-4}$ &  0.18\\
		2   & 0.5      & 1.0  & 40\textdegree & 1.0$\times 10^{-2}$ &  2.00\\
		2   & 1.0      & 2.0  & 40\textdegree & 5.0$\times 10^{-3}$ &  1.00\\
		2   & 1.5      & 3.0  & 40\textdegree & 2.7$\times 10^{-3}$ &  0.54\\ \hline
		1	& 1.0      & 1.0  & 70\textdegree & 4.8$\times 10^{-3}$ &  0.96\\
		2   & 0.25     & 0.5  & 70\textdegree & 1.7$\times 10^{-3}$ &  0.34\\
		2   & 0.5      & 1.0  & 70\textdegree & 1.0$\times 10^{-2}$ &  2.00\\
		2   & 1.0      & 2.0  & 70\textdegree & 5.0$\times 10^{-3}$ &  1.00\\
		2   & 1.5      & 3.0  & 70\textdegree & 2.7$\times 10^{-3}$ &  0.54\\
		
	\end{tabular}
	\caption{An Initially Slow-rotating Uranus}{\textbf{Note.} This table shows the same calculations as in Table \ref{table:odds:1}, but with the planet having an initial spin period of 68.8 hr. $\epsilon_{i}$ is Uranus's initial obliquity. The normalized probability column divides the Probability by 5$\times 10^{-3}$ as in Table \ref{table:odds:1}.}
	\label{table:odds:2}
\end{table}

\begin{figure}[h]
	
	\begin{tabular}[b]{@{}p{0.45\textwidth}@{}}
		
		\includegraphics[width=0.5\textwidth]{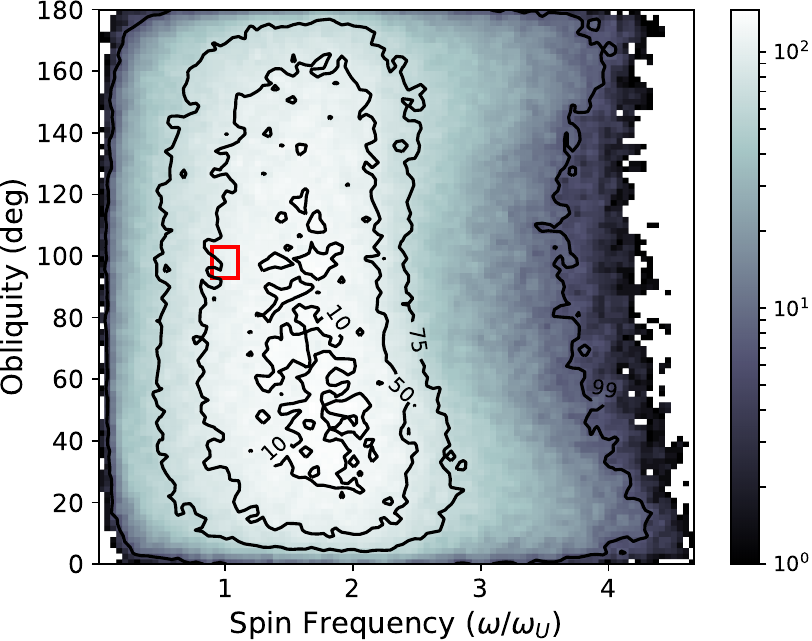}
	\end{tabular}
	
	\caption{Density plot showing two impacts of equal mass ($m_{i}=1.0\,M_{\oplus}$) incident on Uranus with $T_{i}=$68.8 hr and $\epsilon_{i}=$40\textdegree. The probability of Uranus's spin state falling within 10\% of the maximum value is 1.2 times that of the planet's current state ($l_{U}=0.005$).}
	\label{fig:den_slow_GF}
\end{figure}

First, we explore cases where the planet initially spins slowly ($T_{i}\ll 17.2$ hr). In Figure \ref{fig:den_slow_GF} we have Uranus's initial spin period four times slower than its current value, tilted to 40\textdegree, and the planet was struck by two Earth-mass impactors. In this case, even if Uranus was tilted initially by another method, the odds of generating Uranus's current spin state are about the same as if the planet was untilted. This is shown in Table \ref{table:odds:2}, and the entries show similar likelihoods to the nonspinning case. However, both the nonspinning and slow-spinning cases are improbable for two reasons. First, the mechanism responsible for removing excess angular momentum during gas accretion needs to be extremely efficient. And second, the odds that both Uranus and Neptune were spun up similarly by impacts require significant fine-tuning.

\begin{figure}[h]

	\begin{tabular}[b]{@{}p{0.45\textwidth}@{}}
		
		\includegraphics[width=0.5\textwidth]{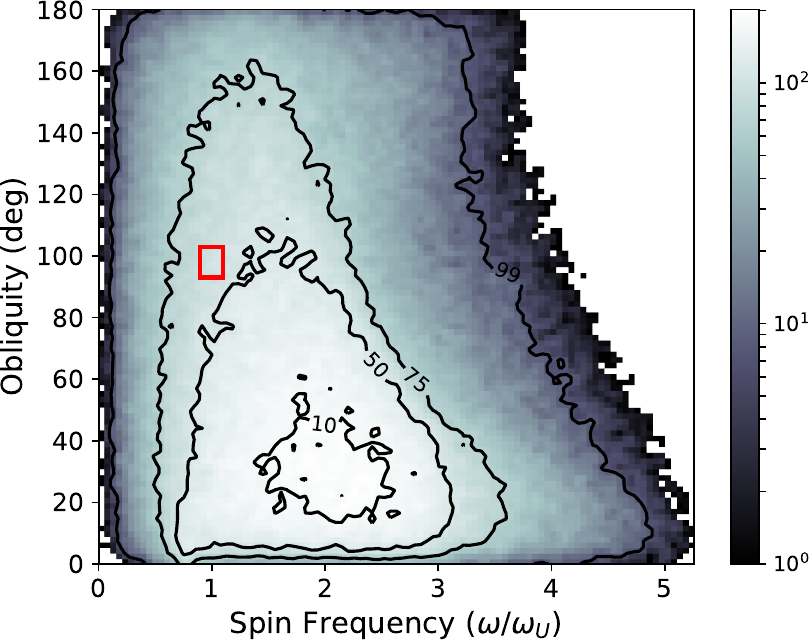}
	\end{tabular}
	\caption{Density plot for collisions incident on Uranus. Two impacts of equal mass ($m_{i}=1.0\,M_{\oplus}$) incident on Uranus with $T_{i}=$17.2 hr and $\epsilon_{i}=$0\textdegree. The color bar shows the number of realizations for that value, and the contour lines contain the values within which a percentage of realizations are found. The red box contains the space within 10\% of Uranus's current obliquity and spin rate. Uranus having a spin of $2\,\omega_{U}$ and $\epsilon=30$\textdegree~is twice as likely as its current state ($l_{U}=0.0042$). }
	\label{fig:den_fast_GF}
\end{figure}

Accordingly, we investigate the effects of gas accretion by considering impacts onto an untilted fast-spinning ($T_{i}= 17.2$ hr) Uranus. Note that because we are adding angular momentum vectors, the order does not matter; therefore, striking Uranus with a giant impactor before the planet accretes gas will yield the same probability distributions as the reverse case considered here. For an initial spin period near Uranus's current value, the minimum impactor mass increases by $\sqrt{2}$ from $\sim0.4\,M_{\oplus}$ to $0.55\,M_{\oplus}$ over the nonspinning case because the planet already has the correct $|\vec{L}|$, which must be rotated by $\sim90$\textdegree~by the impact. However, while the slowly spinning cases have a relatively flat obliquity distribution, a fast-spinning planet is more resistant to change. For example, striking this planet with a 1 $M_{\oplus}$ object will most likely yield little to no change to the planet's spin state (see Figure 7(a) in \cite{2020ApJ...888...60R}). Introducing more impactors does not change this conclusion appreciably; the planet still tends to remain with a low tilt and similar spin period. Figure \ref{fig:den_fast_GF} demonstrates this with the most favorable case of two 1 $M_{\oplus}$ strikes onto an untilted planet already spinning with a 17.2 hr period. Additional cases are reported in Table \ref{table:odds:4}. 

\begin{table}[h]
	\begin{tabular}{l|lcr|l|c}
		N   & $M_{i}$       & $M_{T}$  & $\epsilon_{i}$  & Probability ($l_{U}$) & Normalized Probability  \\ \hline
		1	& 1.0      & 1.0 & 0\textdegree  & 3.4$\times 10^{-3}$ &  0.68\\
		2   & 0.25     & 0.5 & 0\textdegree  & 0 &  0\\
		2   & 0.5      & 1.0 & 0\textdegree  & 3.7$\times 10^{-3}$ &  0.74\\
		2   & 1.0      & 2.0 & 0\textdegree  & 4.1$\times 10^{-3}$ &  0.82\\
		2   & 1.5      & 3.0 & 0\textdegree  & 2.6$\times 10^{-3}$ &  0.52\\  
		5   & 0.6      & 3.0 & 0\textdegree  & 6.1$\times 10^{-3}$ &  1.22\\		
		10  & 0.3      & 3.0 & 0\textdegree  & 7.5$\times 10^{-3}$ &  1.50\\
		15  & 0.2      & 3.0 & 0\textdegree  & 6.0$\times 10^{-3}$ &  1.20\\ \hline
		1	& 1.0      & 1.0 & 40\textdegree & 4.5$\times 10^{-3}$ &  0.90\\
		2   & 0.25     & 0.5 & 40\textdegree & 1.3$\times 10^{-3}$ &  0.26\\
		2   & 0.5      & 1.0 & 40\textdegree & 7.4$\times 10^{-3}$ &  1.48\\
		2   & 1.0      & 2.0 & 40\textdegree & 4.7$\times 10^{-3}$ &  0.94\\
		2   & 1.5      & 3.0 & 40\textdegree & 2.6$\times 10^{-3}$ &  0.52\\ \hline
		1	& 1.0      & 1.0 & 70\textdegree & 8.3$\times 10^{-3}$ &  1.66\\
		2   & 0.25     & 0.5 & 70\textdegree & 2.6$\times 10^{-2}$ &  5.20\\
		2   & 0.5      & 1.0 & 70\textdegree & 1.4$\times 10^{-2}$ &  2.80\\
		2   & 1.0      & 2.0 & 70\textdegree & 5.7$\times 10^{-3}$ &  1.14\\
		2   & 1.5      & 3.0 & 70\textdegree & 2.7$\times 10^{-3}$ &  0.54\\
		
	\end{tabular}
	\caption{An Initially Fast-rotating Uranus}{\textbf{Note.} This table shows the same calculations as in Table \ref{table:odds:1}, but with the planet having an initial spin period of 17.2 hr. $\epsilon_{i}$ is Uranus's initial obliquity. The final column normalizes the probability column by 5$\times 10^{-3}$ as in Table \ref{table:odds:1}.}
	\label{table:odds:4}
\end{table}

\begin{figure}[h]
	
	\centering
	\begin{tabular}[b]{@{}p{0.45\textwidth}@{}}
		\centering
		\includegraphics[width=0.45\textwidth]{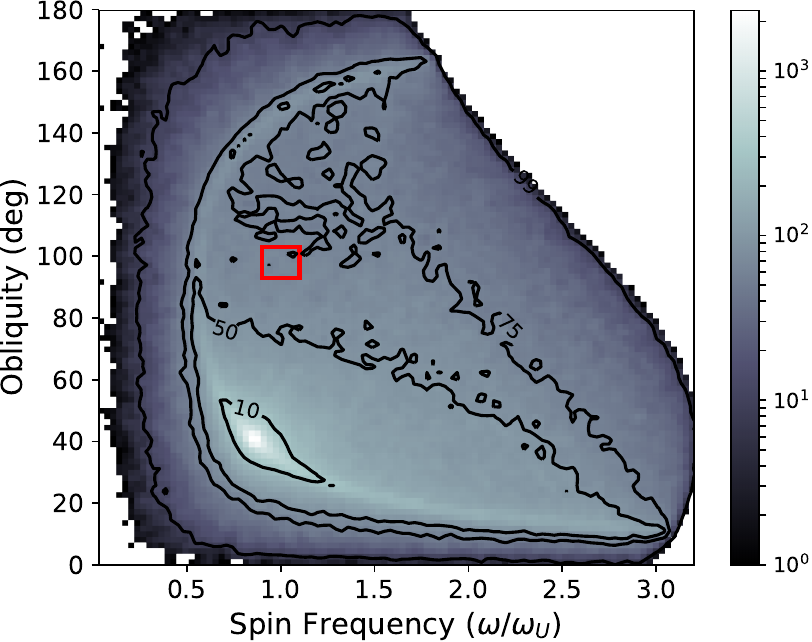}
		\centering (a)
	\end{tabular}\\
	\begin{tabular}[b]{@{}p{0.45\textwidth}@{}}
		\centering
		\includegraphics[width=0.45\textwidth]{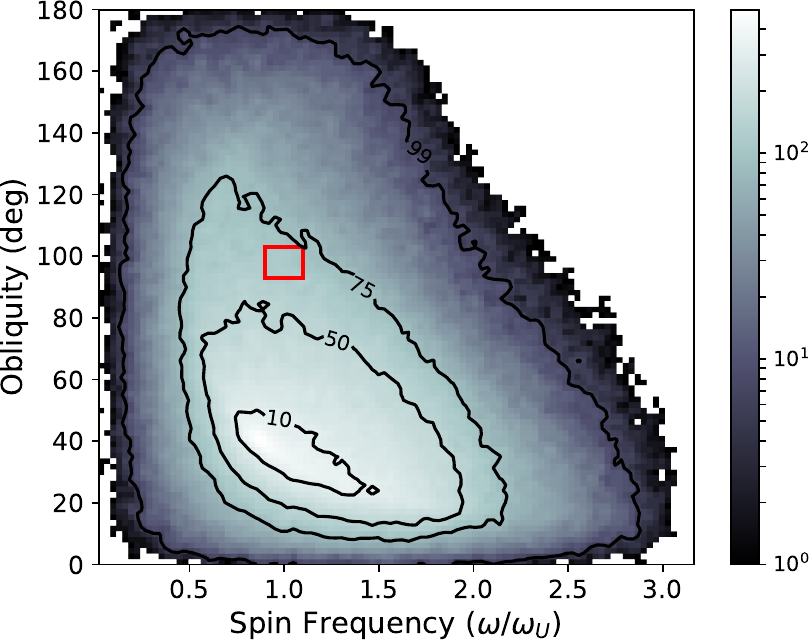}
		\centering (b)
	\end{tabular}\\
	\begin{tabular}[b]{@{}p{0.45\textwidth}@{}}
		\centering
		\includegraphics[width=0.45\textwidth]{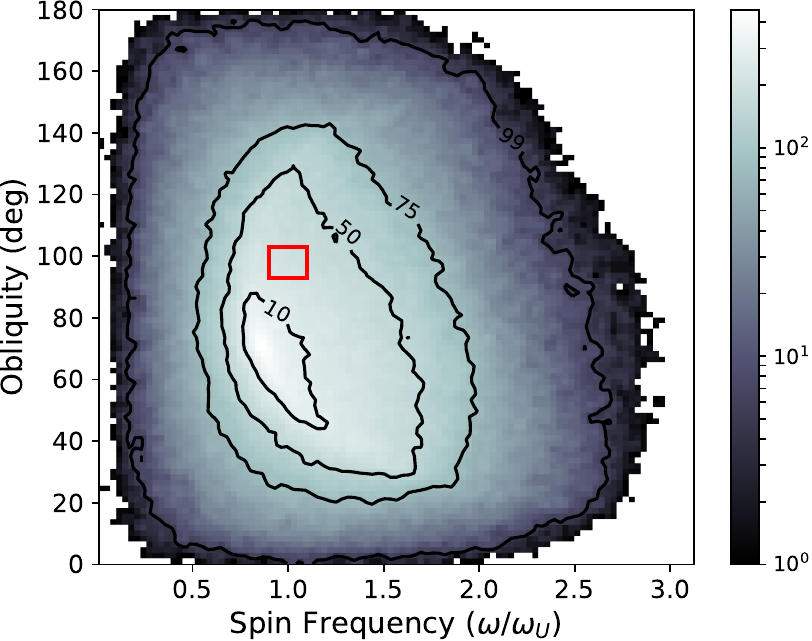}
		\centering (c)
	\end{tabular}
	\caption{(a) Density plot showing one impact ($m_{i}=1.0\,M_{\oplus}$) incident on Uranus with $T_{i}=$17.2 hr and $\epsilon_{i}=$40\textdegree. It is 17.5 times more likely to fall within 10\% of the initial state than Uranus's current spin state ($l_{U}=0.0045$). Notice the sharp spike of over 2000 counts near the planet's initial spin state. (b) Two impacts ($m_{i}=0.5\,M_{\oplus}$) incident on Uranus with $T_{i}=$17.2 hr and $\epsilon_{i}=$40\textdegree. The probability of Uranus's spin state falling within 10\% of the maximum value is 3.5 times that of the planet's current state ($l_{U}=0.0075$). (c) Two impacts ($m_{i}=0.5\,M_{\oplus}$) incident on Uranus with $T_{i}=$17.2 hr and $\epsilon_{i}=$70\textdegree. The probability of Uranus's spin state falling within 10\% of the maximum value is 1.8 times that of the planet's current state ($l_{U}=0.014$).}
	\label{fig:den_fast_40_GF}
\end{figure}

If Uranus was initially tilted by a 40\textdegree~resonance kick, its rapid rotation ensures that its spin state will tend to remain relatively unaffected by subsequent impacts. This can be seen in Figure \ref{fig:den_fast_40_GF}(a) with a 1 $M_{\oplus}$ strike, where the probability of tilting Uranus to 98\textdegree~is only 4.5$\times 10^{-3}$. The odds do improve if the number of impacts increases (Figure \ref{fig:den_fast_40_GF}(b)), but they are not better than the non-spinning case. However, if Uranus was initially tilted by 70\textdegree~via a spin-orbit resonance \citep{2020ApJ...888...60R}, then two 0.5 Earth-mass strikes generate a favorable result (Figure \ref{fig:den_fast_40_GF}(c)). Also, only in this case will two 0.25 $M_{\oplus}$ strikes yield even better likelihoods (see Figure 8 in \cite{2020ApJ...888...60R}). Therefore, if Uranus's and Neptune's current spin rates were a by-product of gas accretion, then a large resonance kick can significantly reduce the mass needed in later impacts.

\begin{table}[h]
	\begin{tabular}{l|lcr|l|c}
		N   & $M_{i}$       & $M_{T}$   & $\epsilon_{i}$   & Probability ($l_{U}$) & Normalized Probability \\ \hline
		1   & 1.0      & 1.0 & 0\textdegree   & 2.3$\times 10^{-3}$  & 0.46\\
		2   & 0.25     & 0.5 & 0\textdegree   & 0  & 0.00\\		
		2   & 0.5      & 1.0 & 0\textdegree   & 2.6$\times 10^{-4}$  & 0.05\\
		2   & 1.0      & 2.0 & 0\textdegree   & 2.7$\times 10^{-3}$  & 0.54\\ 
		2   & 1.5      & 3.0 & 0\textdegree   & 2.0$\times 10^{-3}$  & 0.40\\ \hline
		1   & 1.0      & 1.0 & 40\textdegree  & 4.1$\times 10^{-3}$  & 0.82\\
		2   & 0.25     & 0.5 & 40\textdegree  & 0  & 0\\
		2   & 0.5      & 1.0 & 40\textdegree  & 2.0$\times 10^{-3}$  & 0.40\\
		2   & 1.0      & 2.0 & 40\textdegree  & 4.1$\times 10^{-3}$  & 0.82\\ 
		2   & 1.5      & 3.0 & 40\textdegree  & 2.5$\times 10^{-3}$  & 0.50\\ \hline
		1   & 1.0      & 1.0 & 70\textdegree  & 2.1$\times 10^{-3}$  & 0.42\\
		2   & 0.25     & 0.5 & 70\textdegree  & 1.2$\times 10^{-4}$  & 0.02\\
		2   & 0.5      & 1.0 & 70\textdegree  & 3.3$\times 10^{-3}$  & 0.66\\
		2   & 1.0      & 2.0 & 70\textdegree  & 3.0$\times 10^{-3}$  & 0.60\\ 
		2   & 1.5      & 3.0 & 70\textdegree  & 2.4$\times 10^{-3}$  & 0.48\\ \hline
		5   & 0.8      & 4.0 & 0\textdegree   & 3.4$\times 10^{-3}$  & 0.68\\
		10  & 0.4      & 4.0 & 0\textdegree   & 5.0$\times 10^{-3}$  & 1.00\\
		15  & 0.2667   & 4.0 & 0\textdegree   & 4.4$\times 10^{-3}$  & 0.88\\

	\end{tabular}
	\caption{An Initially Very Fast Rotating Uranus}{This table shows the same calculations as in the previous tables, but the planet is spinning with a period of 8.6 hr. The final column has been normalized as in Table \ref{table:odds:1}.}
	\label{table:odds:3}
\end{table}

\begin{figure}[h]
	
	\begin{tabular}[b]{@{}p{0.45\textwidth}@{}}
		
		\includegraphics[width=0.5\textwidth]{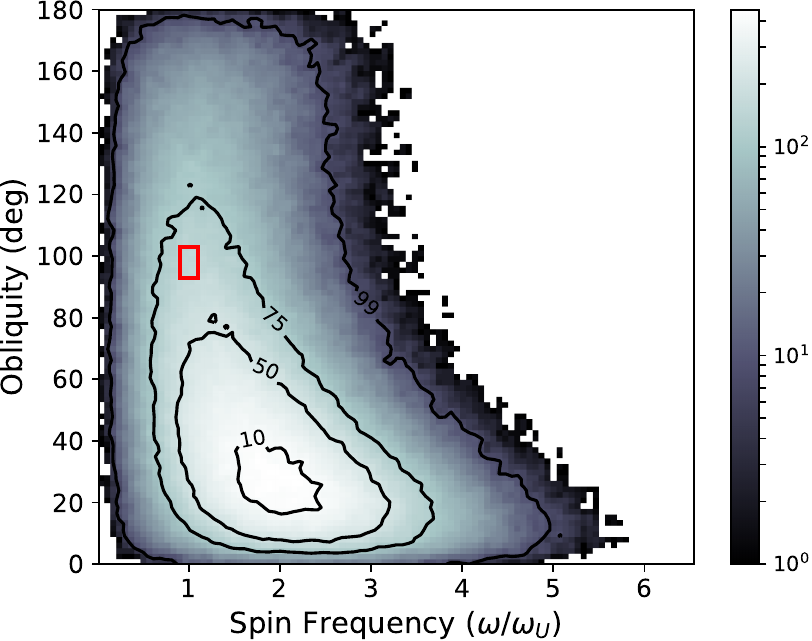}
	\end{tabular}
	
	\caption{Density plot showing 10 impacts of equal mass ($m_{i}=0.4\,M_{\oplus}$) incident on Uranus with $T_{i}=$8.6 hr and $\epsilon_{i}=$0\textdegree. The probability of Uranus's spin state falling within 10\% of the maximum value is 2.9 times higher than falling near the planet's current state ($l_{U}=0.005$), as shown in the red box.}
	\label{fig:10fast_GF}
\end{figure}

Finally, the mechanism that removes angular momentum during gas accretion could have been very weak, and Uranus would have been initially spinning very fast ($T_{i}\gg 17.2$ hr). In this case, slowing down Uranus's spin rate and tilting the planet over would require very massive impacts. As discussed in the previous subsection, changing the planet's spin state with many impactors requires more impacting mass to compensate for the partial cancellations of impact effects. Table \ref{table:odds:3} shows that 10 impacts totaling to $4\,M_{\oplus}$ produce plausible outcomes. While their obliquity distributions peak at around 30\textdegree, which favors a Neptune formation scenario, the planets would still likely be spinning twice as fast as they are today (Figure \ref{fig:10fast_GF}). We would therefore not expect the required massive impactors to spin down both Uranus and Neptune similarly. Additionally, 10 independent strikes is less probable than only 2, while also requiring the solar system to have been populated with many massive rogue planetary cores.

\section{Summary and Conclusion}

We have searched exhaustively for ways to tilt Uranus to 98\textdegree. Because gas accretion provides the giant planets with a significant source of angular momentum \citep{1986Icar...67..391B, 2009Icar..199..338L, 2010AJ....140.1168W}, and the planet's core was likely to have formed from the accumulation of pebbles and planetesimals, any primordial spin state was likely to be erased, leaving near-zero initial obliquities and relatively fast spin rates. As such, changing the planets' obliquities significantly without altering the planet's spin period requires either a specific configuration of large collisions or a secular spin-orbit resonance.

A single Earth-mass strike is certainly capable of reproducing Uranus's spin state, and the debris disk could recreate the present-day satellite system \citep{2020NatAs...4..880I}. If the Uranian satellites were instead formed from the planet's circumplanetary disk, then there needed to have been multiple collisions in order to explain the prograde motion of the Uranian satellites \citep{2012Icar..219..737M}. Maximizing the probability of this outcome requires minimizing both the number of impacts and the mass of each impactor, as there must have been many more rogue Mars-sized cores than Earth-sized ones dispersed throughout the early solar system \citep{2015A&A...582A..99I,2015Natur.524..322L,2015PNAS..11214180L}. We have shown that, in general, two impacts totaling to $1\,M_{\oplus}$ yields the most favorable outcome compared to all the other possibilities, but the odds generally do not change by more than a factor of a few for other scenarios---single impact or otherwise. Furthermore, the likelihood of generating Uranus's current spin state is still very low. An initially fast-spinning planet cannot be tilted easily because of its large initial angular momentum. We could improve the likelihood of generating Uranus's spin state by assuming a slower initial spin period (Figure \ref{fig:den_slow_GF}), but this would require an even more efficient method of removing angular momentum as the planet accretes its gaseous atmosphere; there seems to be little justification for this.

The advantage of the collisionless secular spin-orbit resonance model is that it preserves both Uranus's spin rate and its moons' orbits by gently tipping the Uranian system over. Here we have investigated a resonance argument with Uranus commensurate with Neptune. We have shown that Uranus, being located between Jupiter and Saturn, can augment the planet's spin precession rate enough to match with Neptune located beyond Saturn. Capture into resonance can tilt the planet to near 90\textdegree, but only on unrealistic 100 Myr timescales. Resonance kicks, on the other hand, require just $\sim10^{7}$ yr, but would produce at most a 40\textdegree~obliquity under ideal circumstances. This resonance can, however, easily excite Uranus's obliquity by about 10\textdegree~or 20\textdegree, which would eliminate one of the impacts required by \cite{2012Icar..219..737M}. As we have seen in Tables \ref{table:odds:2} and \ref{table:odds:4}, however, an initial obliquity of 40\textdegree~does not provide much mass reduction or probability improvements in the subsequent collisions needed to generate Uranus's current spin state. We would need to tilt the planet all the way up to $\sim70$\textdegree~to significantly reduce the mass of later impacts, which would be more likely to occur during the time that Uranus harbored a circumplanetary disk \citep{2018ApJ...868L..13S,2020ApJ...888...60R}. Even in ideal circumstances, these noncollisional models cannot drive the planet's obliquity beyond 90\textdegree, and so large collisions seem unavoidable. 

Tilting Uranus is a difficult problem, and each of the models that we have considered contains a major fault. Neptune's 30\textdegree~obliquity, by contrast, can be much more easily explained by any one of these scenarios. Regardless of the planet's initial spin rate, Figures \ref{fig:den_slow_GF} and \ref{fig:den_fast_GF} imply a high probability of generating Neptune's current spin state. If Neptune's spin rate was a by-product of gas accretion, then a small impact or an impact near the planet's center is sufficient to explain Neptune's low obliquity. \cite{2020MNRAS.492.5336R} reinforce this scenario because a head-on collision of a large impactor with Neptune may also explain its core's higher moment of inertia, in opposition to Uranus's more centrally dense interior. Furthermore, if Neptune was instead captured into a spin-orbit resonance, then we require a less massive disk and a smaller orbital inclination than for Uranus to tilt Neptune over \citep{2020ApJ...888...60R}. Because ice giants probably have harbored large circumplanetary disks while accreting their massive atmospheres, then we should expect at least minor obliquity excitations. Under these circumstances, a combination of the two models, a spin-orbit resonance followed by a giant impact, may be the more likely explanation for Uranus's unusual spin state.

\section{Acknowledgement}
This work was supported by NASA Headquarters under the NASA Earth Science and Space Fellowship grant NNX16AP08H. We thank Alexander Dittmann, Pradip Gatkine, and Scott Lawrence for useful discussions, and we thank the anonymous reviewers for helpful feedback. We also thank Leslie Sage for his helpful comments and suggestions on an earlier draft of this manuscript.

\appendix
\section{Angular Momentum and Obliquity Distributions}

Here we derive the angular momentum and obliquity distributions from accreting multiple small particles, similar to the approach of \cite{1993Icar..103...67D}. If these particles are isotropically distributed, then they possess a wide range of eccentricities and inclinations, and so there is no preference for any spin direction. This isotropy breaks down if particles instead orbit within the planetary disk at low inclinations and eccentricities. This discussion draws heavily from \cite{grinstead2006grinstead}.

\subsection{Angular Momentum Distributions}
The calculation for the angular momentum distribution of a planet from multiple strikes at random locations on the planet's surface is a random walk scenario. We start with the magnitude of the spin angular momentum of a planet:

\begin{equation}
L=\sqrt{L_{X}^{2} + L_{Y}^{2} + L_{Z}^{2}}
\end{equation}\label{angmomeq}

\noindent where the probability distribution ($f_{L_{k}}(l_{k})$) of each component ($L_{k}$) of the angular momentum vector is described by a normal distribution as a by-product of the central limit theorem:

\begin{equation}
f_{L_{k}}(l_{k})=\frac{1}{{\sigma}_{k}\sqrt{2\pi}}e^{-{l_{k}^{2}}/{2{\sigma}_{k}^{2}}}.
\end{equation}
As such, to find the distribution of the magnitude of the angular momentum, we will first need to determine the square of each distribution and then the sum of three squares, and finally, to take the square root of the sum as seen in Equation \ref{angmomeq}. 

The distribution of the square of each component ($L_{k}^{2}$) can be calculated by assuming that $X$ and $Y$ are continuous random variables (i.e. ``variates'' as depicted in upper case), with $x$ and $y$ as specific elements in the ranges of their corresponding variates (i.e. also called ``quantiles,'' depicted here in lower case; \citep{grinstead2006grinstead}). $X$ and $Y$ have cumulative distribution functions $F_{X}$ and $F_{Y}$, and $Y$ is described by a strictly increasing function of $X$: $Y=\phi(X)$. $F_{Y}(y)=P(Y \leq y)$, where the right hand side describes the probability that the variate $Y$ is less than or equal to a number $y$, which is equal to $P(\phi(X)\leq y)=P(X\leq \phi^{-1}(y))=F_{X}(\phi^{-1}(y))$. 

So, for the variate $X^{2}$ and its corresponding quantile $x^{2}$:
\begin{equation}
F_{X^{2}}(x^{2})=P(X^{2}\leq x^{2})=P(-x\leq X \leq x).
\end{equation}
The right hand side can be rearranged accordingly:
\begin{equation}
P(-x\leq X \leq x)=P(X\leq x) - P(X\leq -x)
\end{equation}
so that
\begin{equation}
F_{X^{2}}(x^{2})=F_{X}(x)-F_{X}(-x).
\end{equation}

The corresponding density distribution function for an arbitrary variate Y is $f_{Y}(y)=\frac{d}{dy}F_{Y}(y)$. Starting with $F_{Y}(y)=F_{X}(\phi^{-1}(y))$, we take the derivative of each side and employ the chain rule to obtain $f_{Y}(y)=f_{X}(\phi^{-1}(y))\frac{d}{dy}\phi^{-1}(y)$.\\
So,
\begin{equation}
f_{X^{2}}(x^{2})=\frac{f_{X}(x)+f_{X}(-x)}{2x}.
\label{eq3}
\end{equation}

Because the normal distribution is centered at zero and is symmetric, the density distribution for $L_{k}^{2}$ is then

\begin{equation}
f_{L_{k}^{2}}(l^{2})= \frac{1}{{\sigma}_{k}l\sqrt{2{\pi}}}e^{-{l^{2}}/{2{\sigma}_{k}^{2}}}
\label{eq2}
\end{equation}
which is the distribution for a chi squared with one degree of freedom.

Next, the density distribution of the sum of two independent random variables is their convolution. Let $L_{XY}^{2}=L_{x}^{2} + L_{y}^{2}$ and its corresponding density distribution:
\begin{equation}
f_{L_{XY}^{2}}(l_{xy}^{2})=\int_{0}^{l_{xy}^{2}}f_{L_{X}^{2}}(l_{xy}^{2}-l_{y}^{2})f_{L_{Y}^{2}}(l_{y}^{2})dl_{y}^{2}
\label{eq4}
\end{equation}
where $L_{Y}^{2}$ ranges from 0 to $L_{XY}^{2}$. Note that the standard deviations for both $f_{L_{X}}$ and $f_{L_{Y}}$ are equal with $\sigma={\sigma}_{x}={\sigma}_{y}$. Thus, combining Equation \ref{eq2} and \ref{eq4}:

\begin{equation}
f_{L_{XY}^{2}}(l_{xy}^{2})=\frac{1}{2{\pi}{\sigma}^{2}} \int_{0}^{l_{xy}^{2}} \left(e^{-(l_{xy}^{2}-l_{y}^{2})/2{\sigma}^{2}} (l_{xy}^{2}-l_{y}^{2})^{-0.5}\right) \left(e^{-l_{y}^{2}/2{\sigma}^{2}} (l_{y}^{2})^{-0.5}\right) dl_{y}^{2}=\frac{ e^{-l_{xy}^{2}/2{\sigma}^{2}}}{2{\sigma}^{2}}
\end{equation}

Now, let ${L}^{2} = L_{XY}^{2} + L_{Z}^{2}$ and repeat the above process. The probability distribution $f_{{L}^{2}}(l^{2})$ describes the probability that $L^{2}$ takes the value $l^{2}$, and $f_{L_{Z}^{2}}(l_{z}^{2})$ describes the probability that $L_{Z}^{2}$ takes the value $l_{z}^{2}$. We explicitly treat the general case $\sigma\neq{\sigma}_{z}$.

\noindent The density distribution for ${L}^{2}$ is
\begin{equation}
f_{{L}^{2}}(l^{2})=\int_{0}^{l^{2}}f_{L_{XY}^{2}}(l^{2}-l_{z}^{2})f_{L_{Z}^{2}}(l_{z}^{2})dl_{z}^{2}=\frac{1}{2\sqrt{2{\pi}}\,{\sigma}^{2}{\sigma}_{z}} \int_{0}^{l^{2}} e^{-(l^{2}-l_{z}^{2})/2{\sigma}^{2}}e^{-l_{z}^{2}/2{\sigma}_{z}^{2}}\> (l_{z}^{2})^{-0.5}\> dl_{z}^{2}
\end{equation}
let $\beta=\frac{{\sigma}^{2} - {\sigma}_{z}^{2}}{2{\sigma}^{2}{\sigma}_{z}^{2}}$, $\gamma=\beta l_{z}^{2}$, and $d\gamma=\beta dl_{z}^{2}$, and so

\begin{equation}\label{eq:DT}
f_{{L}^{2}}(l^{2})=\frac{e^{-l^{2}/2{\sigma}^{2}}}{2\sqrt{2{\pi}}\,{\sigma}^{2}{\sigma}_{z}}\frac{1}{\sqrt{\beta}} \int_{0}^{\beta l^{2}} e^{-\gamma}\> {\gamma}^{-0.5}\> d\gamma
\end{equation}

Equation \ref{eq:DT} is of similar form to Equation 109 found in \cite{1993Icar..103...67D}. Applying Equation \ref{eq3} to $f_{L^{2}}$ and noting that because $L$ is the magnitude of the planet's angular momentum, $f_{L}(-l)=0$. We find $f_{{L}}(l)=f_{{L}^{2}}(l)\cdot2l$. The probability distribution describing the angular momentum of the planet for $\beta>0$, or ${\sigma}_{x}={\sigma}_{y}>\sigma_{z}$ is then

\begin{equation} \label{eq:angmom1}
f_{{L}}(l) = \frac{l e^{-l^{2}/2{\sigma}^{2}}}{\sqrt{2{\pi}}\,{\sigma}^{2}{\sigma}_{z}}\frac{1}{\sqrt{\beta}}\,\gamma(0.5,\beta l^{2})  
\end{equation}
where $\gamma(0.5,\beta l^{2})$ is the lower incomplete gamma function. For $\beta<0$ (${\sigma}_{x}={\sigma}_{y}<\sigma_{z}$):

\begin{equation}\label{eq:angmom2}
f_{{L}}(l) = \frac{l e^{-l^{2}/2{\sigma}^{2}}}{\sqrt{2{\pi}}\,{\sigma}^{2}{\sigma}_{z}}\frac{1}{\sqrt{-\beta}}(2l\sqrt{-\beta})\,\Phi(0.5;1.5;-\beta l^{2})
\end{equation}
where $\Phi(0.5;1.5;-\beta l^{2})$ is the confluent hypergeometric function of the first kind. For $\beta=0$, where $\sigma={\sigma}_{x}={\sigma}_{y}=\sigma_{z}$ (isotropic case), the form is particularly simple:

\begin{equation}
f_{{L}}(l) = \frac{2l^{2} e^{-l^{2}/2{\sigma}^{2}}}{\sqrt{2{\pi}}\,{\sigma}^{3}}.
\end{equation}

\subsection{Obliquity Distributions}

The obliquity angle ($\epsilon$) is defined by $\tan(\epsilon)=\frac{\sqrt{L_{x}^{2}+L_{y}^{2}}}{L_{z}}=\frac{L_{XY}}{L_{z}}$. To find the distribution of the quotient of two independent variants, we let $Q=X/Y$, where $X$ and $Y$ are independent random variables. Then, $F_{Q}(q)=P(Q\leq q)=P(X/Y\leq q)$. If $Y>0$, then $X\leq yq$, while if $Y<0$, then $X\geq yq$. Therefore, $P(X/Y\leq q) = P(X\leq yq, Y>0)+P(X\geq yq, Y<0)$. These constraints determine the integral limits in the corresponding cumulative distribution:

\begin{equation}
F_{Q}(q)=\int_{y=0}^{\infty}\int_{x=-\infty}^{yq} f_{XY}(x,y)dxdy + \int_{y=-\infty}^{0}\int_{x=yq}^{\infty} f_{XY}(x,y)dxdy.
\end{equation}
and density distribution:
\begin{equation}
f_{Q}(q)=\int_{0}^{\infty} yf_{XY}(yq,y)dy + \int_{-\infty}^{0} (-y)f_{XY}(yq,y)dy.
\end{equation}

\noindent So, to calculate the obliquity distribution, let ${\sigma}\neq{\sigma}_{z}$, and $U=\tan(\epsilon)$ with $u=\tan(\varepsilon)$ as the corresponding quantile. Thus,

\begin{equation}
f_{U}(u)=\int_{0}^{\infty} l_{z} f_{L_{XY}}(ul_{z})f_{L_{z}}(l_{z})dl_{z} + \int_{-\infty}^{0} -l_{z} f_{L_{XY}}(ul_{z})f_{L_{z}}(l_{z})dl_{z}
\end{equation}
which becomes
\begin{equation}
f_{U}(u)=2\int_{0}^{\infty} \frac{|u|{l_{z}^{2}}}{{\sigma}^{2}{\sigma}_{z}\sqrt{2\pi}} e^{-{l_{z}^{2}}u^{2}/(2{\sigma}^{2})} e^{-{l_{z}^{2}}/(2{\sigma}_{z}^{2})} dl_{z}.
\end{equation}

\noindent If we let $\alpha=\frac{u^{2}}{2{\sigma}^{2}} + \frac{1}{2{\sigma}_{z}^{2}}$, then the equation is now of the form
\begin{equation}
\int_{0}^{\infty}t^{2}e^{-\alpha t^{2}}dt = \frac{\sqrt{\pi}}{4{\alpha}^{1.5}}
\end{equation}
and so when normalized:

\begin{equation}
f_{U}(u)=\left|\frac{u}{4\sqrt{2}\,{\sigma}^{2}{\sigma}_{z}{\alpha}^{1.5}}\right|.
\end{equation}
We can change variables to obliquity ($\epsilon$) by setting $f_{\epsilon}(\varepsilon) = \frac{du}{d\varepsilon}f_{U}(u)$ where $\frac{du}{d\varepsilon}=\sec^{2}(\varepsilon)$. We find

\begin{equation}
f_{\epsilon}(\varepsilon)=\left|\frac{1}{4\sqrt{2}\,{\sigma}^{2}{\sigma}_{z}}\frac{\tan(\varepsilon)}{\cos^{2}(\varepsilon)}{{\left(\frac{\tan^{2}(\varepsilon)}{2{\sigma}^{2}} + \frac{1}{2{\sigma}_{z}^{2}}\right)}^{-3/2}}\right|.
\end{equation}
This is equivalent to the obliquity distribution given in \cite[Equation (111)]{1993Icar..103...67D}. For the isotropic case, ${\sigma}_{z}=\sigma$, the distribution reduces to

\begin{equation}
f_{\epsilon}(\varepsilon)=\left|\frac{1}{2}\sin(\varepsilon)\right|.
\end{equation}

\bibliography{bibliography}{}
\end{document}